\documentclass[aps, prl, a4paper,showpacs,twocolumn,10pt]{revtex4-1}
\usepackage{bbm, amsmath, amssymb, amsthm, bm,textcomp, nicefrac,geometry,ragged2e}%, subcaption}
\usepackage{graphicx,epstopdf,color,verbatim,enumitem}
\usepackage[dvipsnames]{xcolor}
\usepackage{float}
\usepackage[bbgreekl]{mathbbol}
\usepackage{graphicx,epstopdf,color,verbatim,enumitem,ulem}

\definecolor{myurlcolor}{rgb}{0,0,0.7}
\definecolor{myrefcolor}{rgb}{0.8,0,0}
\usepackage[unicode=true,pdfusetitle, bookmarks=false,bookmarksnumbered=false,
bookmarksopen=false, breaklinks=false,pdfborder={0 0 0},backref=false,
colorlinks=true, linkcolor=myrefcolor,citecolor=myurlcolor,urlcolor=myurlcolor]
{hyperref}

\usepackage{pbox,hyperref,array}
\usepackage[caption=false]{subfig}

\newcommand{\ket}[1]{\left|#1\right\rangle}

\newcommand{\bra}[1]{\left\langle #1\right|}

\newcommand{\be}{\begin{equation}}
\newcommand{\ee}{\end{equation}}
\definecolor{darkorange}{RGB}{255,140,0}

\newcommand{\bea}{\begin{eqnarray}}
\newcommand{\eea}{\end{eqnarray}}

\makeatletter
\newtheorem*{rep@theorem}{\rep@title}
\newcommand{\newreptheorem}[2]{%
\newenvironment{rep#1}[1]{%
 \def\rep@title{#2 \ref{##1}}%
 \begin{rep@theorem}}%
 {\end{rep@theorem}}}
\makeatother

\newreptheorem{theorem}{Theorem}

\newtheorem*{result*}{Result}

\DeclareGraphicsExtensions{.png,.pdf,.eps}

\begin{document}
\title{Long-range big quantum-data transmission}
\author{M. Zwerger$^{1,2}$, A. Pirker$^1$, V. Dunjko$^{1,3}$, H. J. Briegel$^{1,4}$ and W.~D\"ur$^1$}
\affiliation{$^1$ Institut f\"ur Theoretische Physik, Universit\"at Innsbruck, Technikerstra{\ss}e 21a, 6020 Innsbruck, Austria \\
$^2$ Departement Physik, Universit\"at Basel, Klingelbergstra{\ss}e 82, 4056 Basel, Switzerland \\
$^3$ Max-Planck-Institut f\"ur Quantenoptik, Hans-Kopfermann-Stra{\ss}e 1, 85748 Garching, Germany \\
$^4$ \mbox{Fachbereich Philosophie, Universit\"at Konstanz, Universit\"atsstra{\ss}e 10, 78464 Konstanz, Germany}}
\date{\today}

\begin{abstract}
We introduce an alternative type of quantum repeater for long-range quantum communication with improved scaling with the distance. We show that by employing hashing, a deterministic entanglement distillation protocol with one-way communication, one obtains a scalable scheme that allows one to reach arbitrary distances, with constant overhead in resources per repeater station, and ultrahigh rates. In practical terms, we show that also with moderate resources of a few hundred qubits at each repeater station, one can reach intercontinental distances. At the same time, a measurement-based implementation allows one to tolerate high loss, but also operational and memory errors of the order of several percent per qubit. This opens the way for long-distance communication of big quantum data.
\end{abstract}
\pacs{03.67.Hk, 03.67.Lx, 03.67.-a}
\maketitle

%%%%%%%%%%%%%%%%%%%%%%%%%INTRODUCTION%%%%%%%%%%%%%%%%%%%%%%%%%%%
\paragraph{Introduction.---}
Long-range quantum communication is a prominent application of emerging quantum technologies. It is a building block of quantum networks, with applications to secure channels \cite{Aschauer02,SecChannelPortmann,SecChannelGarg,SecChannelBroadbent,Pirker16a}, distributed quantum computation \cite{CiracDistributed,Meter06,Meter08,Beals20120686} or distributed sensing \cite{Komar2014,Eldredge2016}.
Despite the quantum mechanical limits of repeater-less distribution of quantum information \cite{Wo82,Pir2017}, schemes which achieve the transmission of quantum information over noisy channels have been suggested.
One approach uses quantum error correction (QEC),  performed at regularly spaced stations, to protect quantum information \cite{Knill96,Zw14,Muralidharan2014,Loock16}. Here the transmission is fast, however error thresholds for channel noise and local operations are rather stringent. Additionally, the overhead, i.e., the number of qubits that need to be processed and stored locally, are substantial, growing polylogarithmically with the distance. Entanglement-based quantum repeaters \cite{Br98} (see also \cite{ladd2006,Loock2006,Childress2006,Hartmann2007,Jiang2007,Collins2007,Jiang2009,Bratzik13,Azuma2015}) present a viable alternative, where entanglement is distributed over short distances, and a (nested) combination of entanglement swapping and distillation is used to create high fidelity entangled pairs over longer distances. Using recurrence-type entanglement distillation with two-way classical communication \cite{Be96a,De96}, one obtains a scalable scheme with high noise tolerance for the channel and local operations, polynomially growing local resources and moderate rates \cite{Br98}. The latter are mainly caused by the classical communication waiting times in entanglement distillation and can be overcome by using entanglement distillation protocols (EDP) with one-way communication \cite{Hartmann2007}.

Here, we present an alternative entanglement-based quantum repeater scheme utilizing hashing \cite{Be96,Zw14H} -- an efficient, deterministic EDP with one-way classical communication.
This allows the replacing of the nested entanglement purification and swapping of schemes based on recurrence protocols by a non-nested scheme, leading to an improved scaling of the required local resources with the distance \footnote{Simply using hashing instead of recurrence-based purification in standard schemes still yields a polynomial scaling. Critically, here we employ a deterministic one-way EDP with non-zero yield and the non-nested setting to achieve the constant scaling.}.
Our scheme can handle channel errors and loss as well as operational and memory errors. It features ultra-high rates and large error thresholds achieved by a measurement-based implementation \cite{Zw12,Zw13,Zw14,Zw14H,Zw15}. One-way classical communication also minimizes the required memory time, thereby reducing possible sources of imperfections. More importantly, the overhead in local resources, i.e., the number of ancillary qubits and operations needed at each repeater station per final qubit, is constant, i.e., independent from the distance. This is in stark contrast to previous schemes, where local resources grow polylogarithmically, or even polynomially.
Furthermore, one can combine this approach with a heralded scheme to deal with arbitrary channel loss, the dominant source of noise in fiber or free-space photon transmission. This paves way towards efficient long-distance big quantum-data transmission, the essential ingredient in future quantum networks \cite{Kimble08}.

\paragraph{Setting and scheme.---}
We consider the settings where the quantum channel and the local processing of quantum information are lossy and/or noisy. To circumvent the problem of the absorption probability of the channel (e.g. optical fiber connecting repeater stations) growing exponentially quickly in the distance, we divide the channel into $N$ segments of length $l_0=L/N$, over which (noisy) Bell pairs are generated. One can also use heralded schemes to handle arbitrary (non-unit) channel loss. We assume $n$ such Bell pairs are generated over each segment using $n_c$ parallel channels. 
The noisy Bell pairs between two neighboring nodes are purified using the hashing EDP \cite{Be96},  deterministically generating a fraction of $cn$ output pairs, where $c$ depends on the initial pairs entropy.
The resulting pairs are connected at the intermediate nodes via entanglement swapping, thereby generating $cn$ long-distance entangled pairs between the end nodes. Given perfect local operations, hashing produces ideal pairs (asymptotically in $n$), that can be used to yield perfect long-distance entangled pairs. Below we show how a measurement-based implementation \cite{Zw12,Zw14H} allows us to obtain a scheme generating entangled Bell pairs over arbitrary distances in the imperfect setting, where only the end node noise limits the fidelity. All operations are parallelizable, as only one-way classical communication is required, and all Pauli correction operations, occurring in the protocol, can be postponed to be performed just at the final outputs. The overall scheme is summarized in Fig. \ref{Figure_Setup}. A purely QEC-based version without local two-way communication is also conceivable (see appendix).

\begin{figure}[ht]
\centering
\includegraphics[scale=0.35]{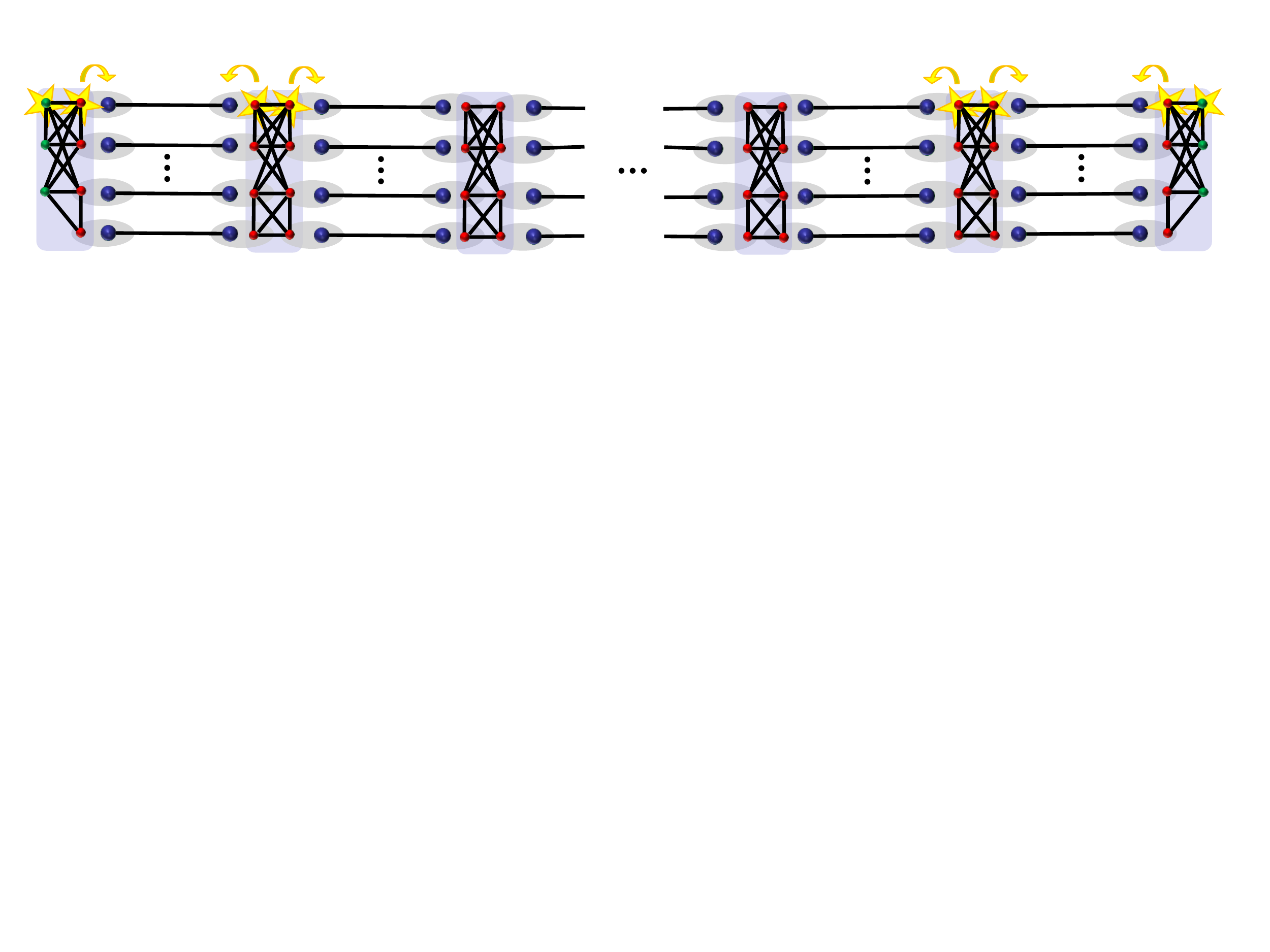}
\vspace{-50mm}
\caption{Illustration of a quantum repeater based on hashing. The channel is divided into $N$ elementary segments, where short-distance entangled pairs are generated over all segments, i.e., between all repeater stations, in parallel. Entanglement distillation via hashing and entanglement swapping are performed in a measurement-based way, by coupling the elementary pairs via Bell measurements to the locally stored resource state. In contrast to quantum repeaters based on recurrence protocols, no nesting is required. Direct encoded transmission would consist in sending encoded information sequentially through the channel.  Please note that this is only an illustration, the real resource states contain at least order of one hundred qubits.}
\label{Figure_Setup}
\end{figure}

\paragraph{Measurement-based hashing.---}
We now briefly describe the key elements of our scheme, hashing and its measurement-based implementation, and discuss their features ensuring the efficiency and functionality in noisy settings.

Hashing distillation protocols operate collectively on a large ensemble of $n$ noisy Bell-pairs. In a single round, bilateral CNOT operations between a subset of ${\mathcal{O}}(n)$ pairs and a target pair are applied, and the target pair is measured. This reveals information about the remaining ensemble, thereby purifying it. Repeating such rounds generates a fraction $cn$ of perfect pairs deterministically in the limit $n \to \infty$. The protocol thus has a non-zero yield $c$ in the noiseless case and only requires one-way classical communication. However, standard hashing fails if operations are noisy. As ${\mathcal{O}}(n)$ operations act on a single qubit, noise accumulates, washing out all information \cite{Zw14H}. We resolve this using a measurement-based implementation \cite{Zw14H}, where local noise up to $7\%$ per qubit, for imperfect resource states and imperfect measurements, is tolerated.

In a measurement-based implementation, quantum information is processed by measurements rather than gates \cite{Ra2001,Br2009}. Similarly to teleportation, input qubits are coupled to an entangled resource state via Bell measurements, realizing the desired operation. For operations that include only Clifford gates and Pauli measurements -- which is the case for EDP and entanglement swapping protocols considered here-- the procedure is deterministic and the resource state consists of only input and output qubits. In fact, qubits that are measured in the Pauli basis (e.g., the target pairs in the hashing protocol) are unnecessary -- a modified, smaller, resource state suffices, where the measurement results can be deduced from the in-coupling Bell measurement outcomes. The resource state corresponding to the hashing protocol has $n$ input and $cn$ output qubits, as the hashing protocol maps $n$ Bell pairs to $cn$ final pairs. The resource state at intermediate repeater stations, which combines hashing and entanglement swapping, is of size $2n$ (there are no output qubits, as entanglement swapping is performed on $cn$ output pairs of the hashing protocol).
This principle was used in \cite{Zw12,Zw15} to obtain resource states of minimal size for a recurrence-based repeater, and in \cite{Zw15,Pirker16} the explicit construction of resource states for different tasks is considered. The key feature, that even complex circuits with many gates, can be implemented with a small resource state (in particular excluding qubits that are measured at any stage of the protocol) leads to a remarkable robustness of measurement-based implementations \cite{Zw12,Zw13,Zw14,Zw14H,Zw15}.

In a measurement-based approach, the noise is manifest in imperfect resource states and Bell measurements. We assume a local noise model for the resource states where local depolarizing noise (LDN) is applied independently to each of the resource qubits (see also the appendix), as in \cite{Zw12,Zw13,Zw14,Zw14H,Zw15}. Such a model is faithful if resource states are affected by local decoherence, or are themselves generated via distillation, as explained in \cite{Raussendorf2006} and \cite{Wallnofer16}. Furthermore, this model accounts for the fact that generating entangled states of a larger number of qubits is experimentally more demanding. The imperfect Bell measurements are also modeled by local noise preceding an otherwise perfect measurement. Memory errors, modeled by local depolarizing noise, can also be accounted for in this way.

When performing a Bell measurement, one can effectively shift the noise between the two qubits \cite{Zw13,Zw15}. In particular, one can (formally) move the noise from input qubits of the local resource states onto the input Bell pair qubits, see figure \ref{Figure_Setup}, resulting in perfect resource states. Only noise on output qubits needs to be considered, which can be done afterwards. Hence, a noisy protocol is equivalent to a {\it perfect} protocol acting on more noisy inputs, where the output state is subsequently affected by local noise.

\paragraph{Repeater scheme in asymptotic noisy setting.---}
We now apply these insights to our repeater protocol in a  setting where channels are lossy and noisy, entanglement distillation and Bell measurements are imperfect and memory errors for the storage of resource states or entangled pairs are accounted for. All noise processes can be included in noise acting on resource states, as argued above (for details regarding memory errors see appendix).

Resource states that we use at intermediate repeater stations have only input qubits, hence all noise can be (formally) moved to input pairs. Thus perfect hashing followed by perfect entanglement swapping is performed on more noisy Bell pairs. As perfect hashing asymptotically produces perfect states, we are in a situation where {\it perfect} Bell states are connected via entanglement swapping. This leads to Bell states at the end nodes, which are affected only by one-step local noise at the final stations. Note that the noise that acts at these final stations is independent from the distance, and is the only factor which determines the final achievable fidelity, in an asymptotic setting. The error threshold for the overall repeater scheme is the same as for measurement-based hashing, up to $7\%$ local noise per qubit.

\paragraph{Communication rates and multiplexing.---}
Our version of the hashing protocol operates on $n$ initial pairs, generated over short distance with sufficiently high fidelity. For instance, one can use a probabilistic (but heralded) scheme at this stage, where a pair is generated with probability $\eta$. We denote the required time that involves pair creation, photon transmission, classical communication time for heralding within an elementary segment by $t_0$. $\eta$ includes channel loss and probabilistic interfaces, and can in principle be arbitrary small. The time required for the local processing of the pairs (in our case, the time to perform the Bell measurements) is denoted by $t_p$. 
In order to minimize the waiting time (and maximize the rate), we use $n_c$ parallel channels. Choosing $n_c = n\left(1/\eta + \epsilon\right)$ suffices to obtain an elementary pair on $n$ of these channels, except with probability ${\mathcal{O}}(e^{-\epsilon^2n})$, from which $m=cn$ long-distance pairs are deterministically generated. We can choose $\epsilon = n^{-1/4}$, such that it vanishes as $n$ increases. We obtain $m$ Bell pairs over all $N$ links within a {\it single} time step $t_0$ with exponentially increasing probability $\left(1-{\mathcal{O}}(e^{-\epsilon^2n})\right)^N$. Only the classical communication time $t_c=L/c_{fiber}$ ($c_{fiber}$ is the speed of light in fiber) to transmit measurement outcomes depends on the distance $L$. The rate per channel is then given by $R = \tfrac{c\eta}{t_0 + t_p}$ in the limit $n \to \infty$. The classical communication time $t_c$ does not enter because one can already start to process new elementary Bell pairs once the pairs from the previous round are processed. Note that $t_0$ can be made as small as the processing time by making the elementary segments short enough. The rate $R$ is thus ultimately limited by $\tfrac{c\eta}{t_p}$, and thus by $t_p$, which is also the time scale which limits the rate of QEC-based repeaters \cite{Knill96}. For more details and examples see appendix.

\paragraph{Hashing and repeaters with finite number of copies $n$.---}
So far we considered the scaling properties of the protocol in an asymptotic setting. Next, we show that for any fixed channel length, a finite number of pairs suffices. For this, we bound the fidelity of the resulting Bell pairs from the basic hashing from below. With this, one can then compute the fidelity of the final Bell pairs resulting from our protocol, the required number of copies for a hashing-based repeater, and the overall efficiency.
Hashing produces $m=cn$ resulting Bell pairs out of $n$ initial/noisy Bell pairs, which is also the number of final, long-distance output pairs, as hashing is deterministic. The yield is given by $c=m/n=1-S(W)-2\delta$ \cite{Be96}, where $S(W)$ is the entropy of the ensemble of initial pairs and $\delta$ is a parameter which affects both the yield and the fidelity for finite sizes. The overhead per pair at each repeater station is determined by $O=4n/m$ as $2n$ qubits are needed for the resource state and another $2n$ for the Bell pairs. The overhead is thus given by $O = 4(1-S(W)-2\delta)^{-1}$ and reaches the constant $4(1-S(W))^{-1}$, which does not scale with the distance $L\sim N$, in the large $n$ limit.

Next, we compute how the distance affects the final pair fidelity, before the noise of the local devices acts on the output pairs at the final repeater stations. This quantity, called {\it{private fidelity}}, bounds the correlations which an eavesdropper might have with the output pairs given the last noise step is independent of the eavesdropper \cite{Aschauer02,Pirker16a,Zwerger17}. Due to the measurement-based implementation we only need to analyze the scaling of the noiseless setting. The hashing protocol succeeds with a probability of $1-\mathcal{O}(\text{exp}(-n \delta^2))$ \cite{Be96}, provided that the fidelity of the initial pairs is large enough (for Werner states the minimum fidelity is $F_{min} \approx 0.8107$). An appropriate choice of $\delta$, such as $\delta =n^{-1/4}$, ensures that the success probability approaches unity. For the quantum repeater to succeed, the entanglement distillation processes at each of the $N$ segments have succeed. The number of links $N$ is proportional to the total length of the channel. For the global, private fidelity of all $m$ outputs, one then obtains
\begin{align}
F_{\rm{gp}} \ge (1 - \alpha \text{exp}(-\beta n \delta^2 ))^N \approx 1 - N \alpha \text{exp}(-\beta n \delta^2) \label{eq:nlinks}
\end{align}
where $\alpha$ and $\beta$ are constants depending on the form of the input Bell pairs (see also the appendix). This shows that the choice of the number $n$ of initial pairs has to depend on $N$, and therefore the length. While this number is increasing, the overhead per transmitted qubit is constant. Choosing $n$ such that $N \alpha \text{exp}(-\beta n^{1/2}) < \epsilon$ with $\epsilon$ small leads to $F_{\rm{gp}}$ close to unity, i.e., $F_{\rm{gp}}\geq1-\epsilon$.
We note that, from a practical perspective, one would however like to limit $n$, as a resource state of size $2n$ needs to be stored at each repeater station.
The fidelity in eq. \ref{eq:nlinks} is the fidelity of the entire set of $m$ output pairs relative to a tensor-product state of $m$ perfect pairs, and consequently,  the same value is a (lousy) bound for the final fidelity of the individual pairs.
From this one can also compute (a bound on) the output fidelity by applying the local depolarizing noise from the output qubits of the resource states.

\begin{figure}
 \centering
 \subfloat[\centering]{\includegraphics[width=0.48\columnwidth]{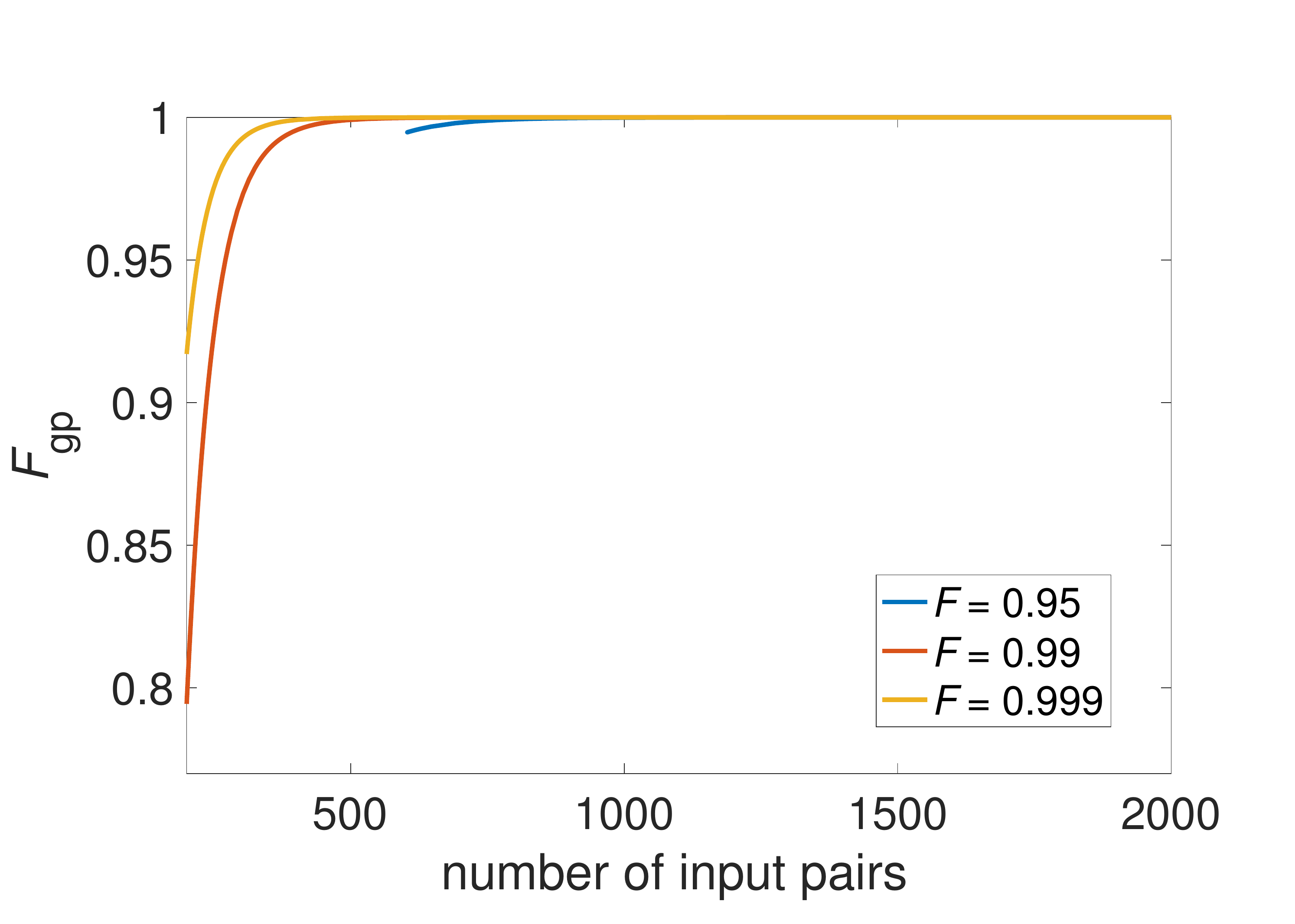} \label{fig:connect}}
 \hfill
 \subfloat[\centering]{\includegraphics[width=0.48\columnwidth]{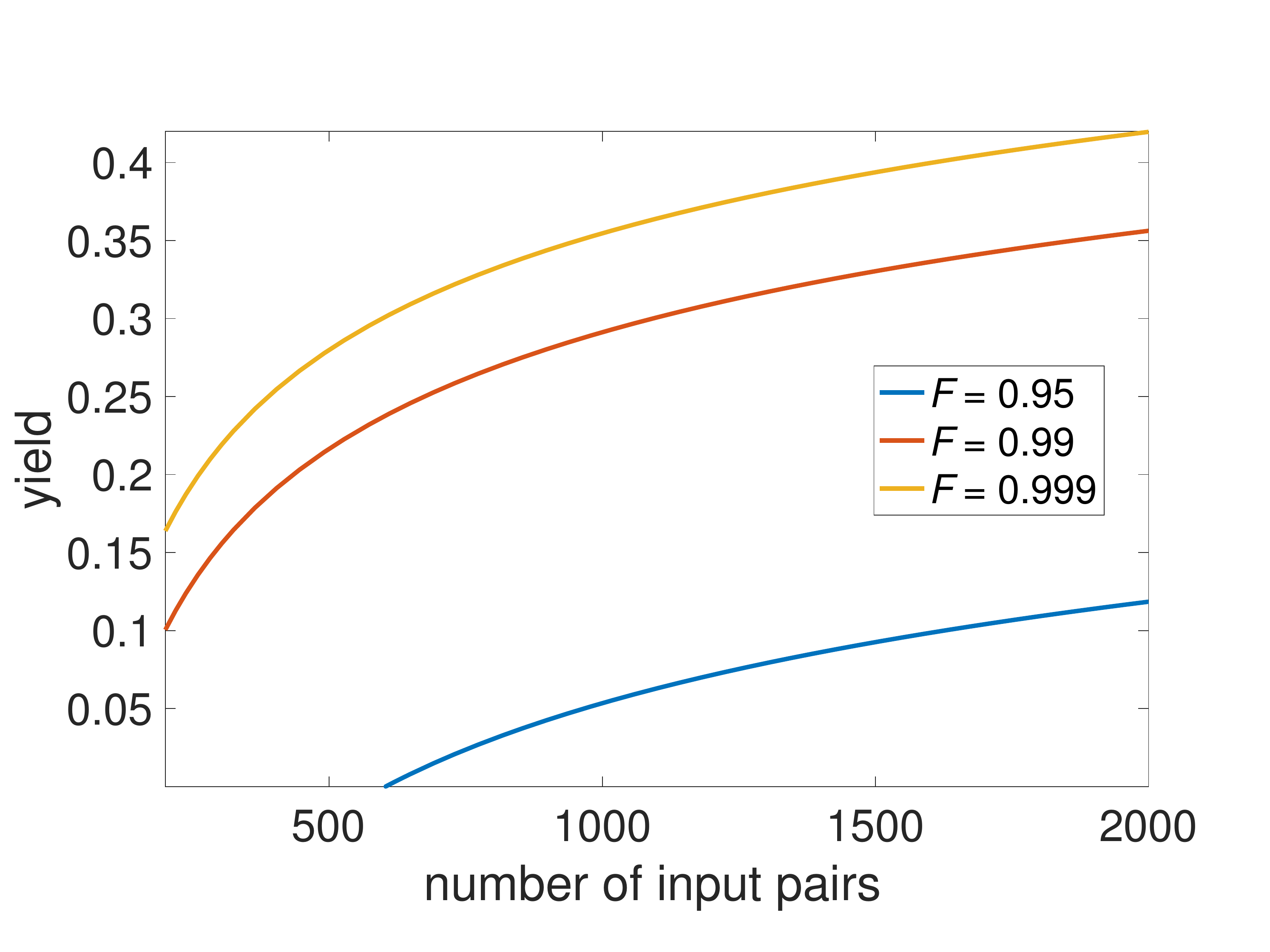} \label{fig:illu_layers}}
 \vspace{-3mm}
 \centering
 \subfloat[\centering]{\includegraphics[width=0.48\columnwidth]{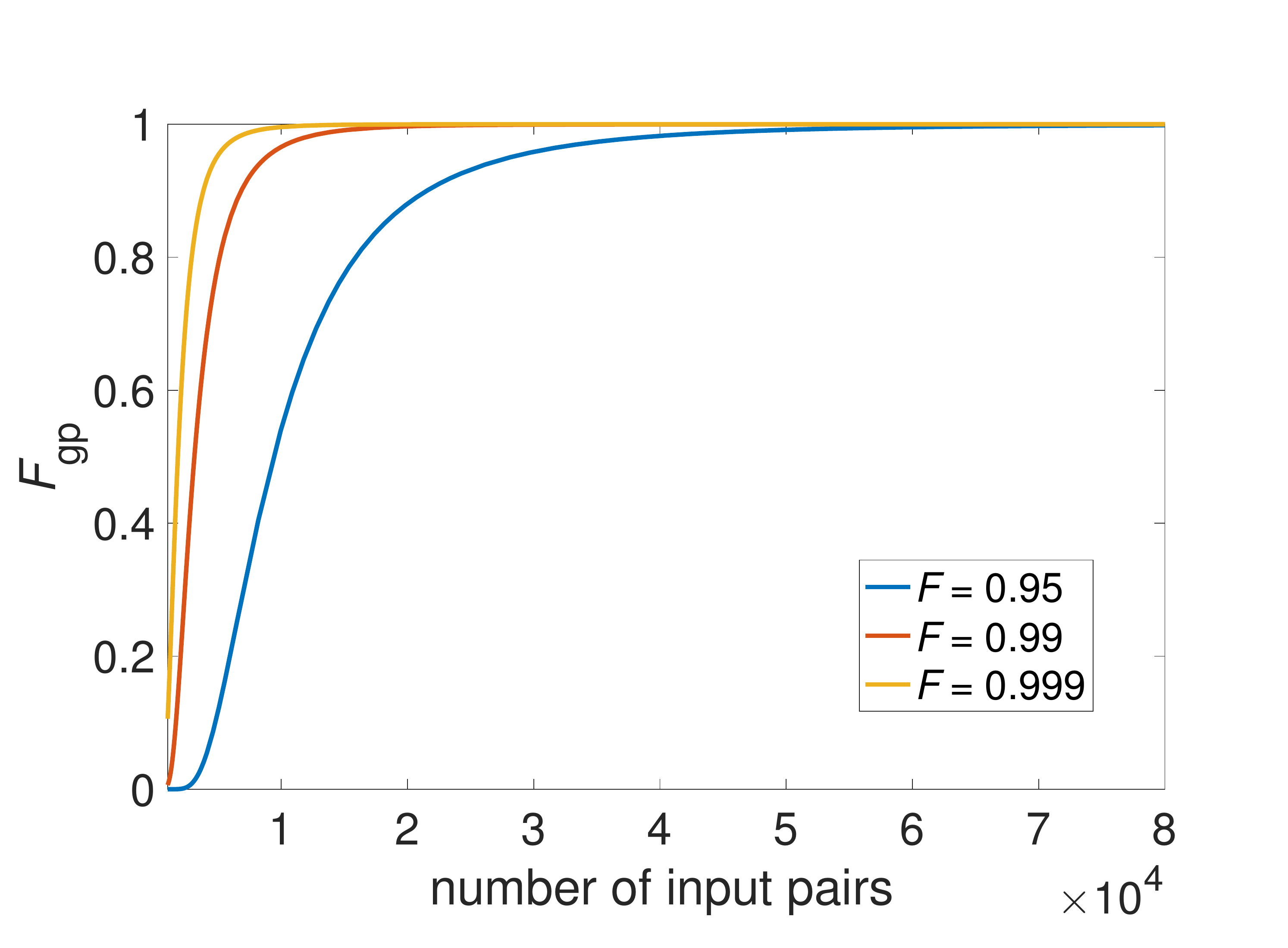} \label{fig:connect}}
 \hfill
 \subfloat[\centering]{\includegraphics[width=0.48\columnwidth]{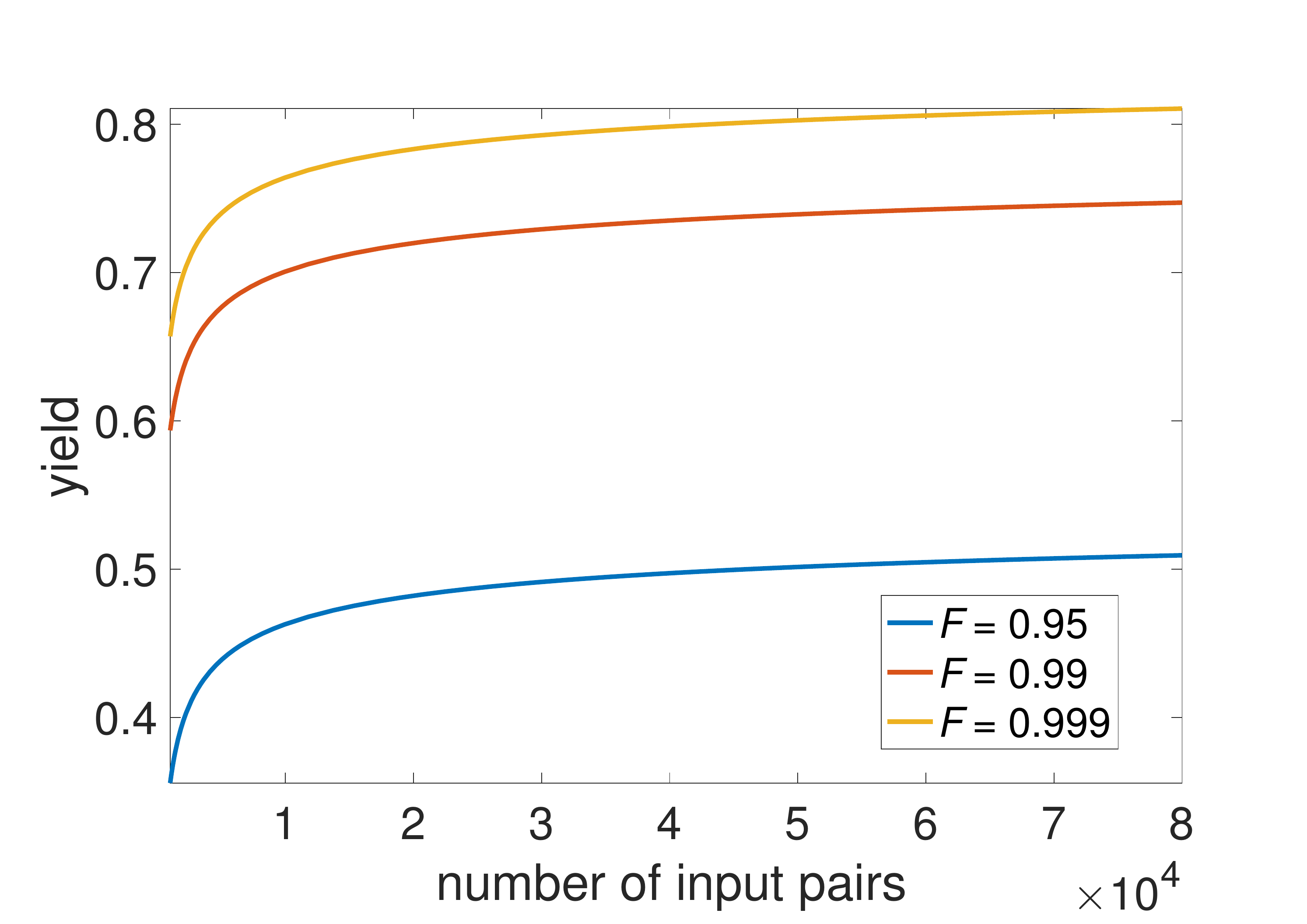} \label{fig:illu_layers}}
 \caption{Plot of the global, private fidelity and yield as a function of the number of initial pairs for $\delta=n^{-1/5}$ (a,b) and $\delta=n^{-1/3}$ (c,d). $F$ denotes the fidelity of the initial Bell pairs, the number of repeater links is $N=100$. We assume local depolarizing noise of $1\%$ per qubit. The fact that the blue curve in (a) seems to starts ``out of the blue" at around $n\approx 600$ is a consequence of the vanishing yield below this number (see (b)). In the choice of $\delta$ there is tradeoff between a higher fidelity (larger $\delta$) and a higher yield (smaller $\delta$). Additional data for more links can be found in the appendix.}
 \label{FigF_Yield}
 \vspace{-3mm}
\end{figure}

For an illustration of the bounds on the global, private fidelity and the yield $c$ for different values of the fidelity of the initial pairs for reasonable parameters, see Fig. \ref{FigF_Yield}.

We obtain the highest attainable fidelity if one measures all initial pairs except one, leading to a $n \to 1$ hashing protocol. The performance of the $n \to 1$ protocol is discussed in detail in the appendix. The required number of copies to achieve purification depends on the initial fidelity of the pairs, where for channel noise of several percent a few hundred copies suffice.

\paragraph{Comparison of approaches}
The main advantage of our scheme over existing ones \cite{Knill96,Br98,Hartmann2007,Jiang2009} is the superior scaling of the local resources with the distance, which is reduced from polynomial \cite{Br98,Hartmann2007} or polylogarithmic \cite{Knill96,Jiang2009} to constant. The robustness to operational errors is comparable for all approaches assuming a measurement-based implementation \cite{Zw12,Zw14,Zw14H}. Our scheme shares the high tolerance of loss errors during transmission with other entanglement-based quantum repeater architectures \cite{Br98,Hartmann2007,Jiang2009}, which is due to the fact that one can use heralded schemes to create the initial Bell pairs. QEC-based schemes \cite{Knill96} are constrained, with a fundamental limit of 50\% loss tolerance imposed by the no-cloning theorem \cite{Wo82}. The long distribution times of the 1998 protocol \cite{Br98} are avoided since hashing is a deterministic one-way EDP. For a comparison of key features of quantum repeater protocols see Table \ref{tablecomp}. In the appendix we also compare the achievable rates and fidelities for our, and the 1998 protocol \cite{Br98} for a measurement-based implementation with $1\%$ LDN, up to $10^4$ links. We find that the rates are up to nine orders of magnitude higher, and anticipate that they are two to three orders of magnitude higher compared to what QEC based quantum repeaters \cite{Knill96} achieve. Thus our new scheme, beyond superior asymptotic performance, also yields better numbers in real world regimes.

\begin{widetext}

\begin{table}[t]
\caption{Comparison of key features of different quantum repeater architectures \cite{Knill96,Br98,Hartmann2007,Jiang2009} and our new protocol.}
\vspace{0.8cm}
\centering
\begin{tabular}{| c | c | c | c | c | c |}
\hline
scheme & Knill \& Laflamme & \parbox[c]{2.8cm}{Briegel, D\"ur, \\ Cirac \& Zoller} & \parbox[c]{3cm}{Hartmann, Kraus,\\  Briegel \& D\"ur} & \parbox[c]{3cm}{Jiang, Taylor,\\ Nemoto, Munro,\\ Van Meter \& Lukin} & \parbox[c]{3cm}{Zwerger, Pirker,\\ Dunjko, Briegel\\ \& D\"ur } \\[3ex]
\hline
year & 1996 & 1998 & 2007 & 2009 & 2017 \\
\hline
based on & QEC & \parbox[c]{2.8cm}{Bell pairs \&\\ two-way EDP} & \parbox[c]{3cm}{Bell pairs \& \\ one-way EDP} & \parbox[c]{3cm}{Bell pairs \& \\ QEC} &  \parbox[c]{3cm}{Bell pairs \& \\ hashing} \\[2ex]
\hline
\parbox[c]{2.2cm}{scaling of \\ local resources} & $\mathcal{O}\left(\text{polylog}(L)\right)$ & $\mathcal{O}\left(\text{poly}(L)\right)$ & $\mathcal{O}\left(\text{poly}(L)\right)$ & $\mathcal{O}\left(\text{polylog}(L)\right)$ & constant \\[2ex]
\hline
\parbox[c]{2.4cm}{rate \\ determined by} & \parbox[c]{2.8cm}{ $\frac{1}{\text{polylog}(L)\cdot t_p}$} & \parbox[c]{2.8cm}{$\frac{1}{\text{poly}(L)\cdot t_c}$} & \parbox[c]{3cm}{$\frac{1}{\text{poly}(L)\cdot max(t_p,t_0)}$} & \parbox[c]{3cm}{ $\frac{1}{\text{polylog}(L)\cdot max(t_p,t_0)}$} & \parbox[c]{3cm}{$\frac{1}{\text{constant}\cdot max(t_p,t_0)}$} \\[1ex]
\hline
\parbox[c]{2.4cm}{constraint \\ on loss} & \parbox[c]{2.8cm}{yes} & \parbox[c]{2.8cm}{no} & \parbox[c]{3cm}{no} & \parbox[c]{3cm}{no} & \parbox[c]{3cm}{no} \\[1ex]
\hline
\end{tabular}
\label{tablecomp}
\end{table}

\end{widetext}

We note that since hashing protocols for the distillation of general graph states exist as well \cite{Du07}, the extension of our architecture to general multipartite quantum networks \cite{Wallnofer16_2D} is straightforward.

\paragraph{Summary and conclusion.---}
We have constructed a quantum repeater which operates with a constant local overhead. This is in stark contrast to all previous long-range communication proposals, which exhibit polynomial or poly-logarithmical overheads in local resources. This guarantees a non-zero yield, high rates and error thresholds for resource states of several percent, and opens the way for big data long-distance quantum communication. The scheme requires only short-time quantum memories for large resource states, and even intercontinental distances can be reached using only a few hundred qubits storage at each repeater station. The protocol has a computational overhead -- the determination of the local correction operations from the classical hash functions, which is generally computationally expensive and might become relevant when the number of pairs becomes very large \cite{Lo09}.
Even this eventuality could be circumvented by either using concatenated hashing of moderate-sized blocks, as discussed above, or through different one-way entanglement distillation protocols (with the same key features as hashing), based on e.g. efficiently decodable low-density parity check codes \cite{Lo09,Gottesman13} or Polar codes \cite{Renes12}.

Our approach requires short-time storage of a number of qubits at each repeater station which is, arguably, large when compared to recent works focused on readily implementable settings. However, our scheme compensates by overcoming many of the drawbacks of existing schemes: it achieves high rates, makes repeaters fully scalable with a small overhead, while being robust against realistic channel and memory errors, and loss.

\paragraph{Acknowledgements.---} This work was supported by the Austrian Science Fund (FWF): P28000-N27 and SFB F40-FoQus F4012, by the Swiss National Science Foundation (SNSF) through Grant number PP00P2-150579, the Army Research Laboratory Center for Distributed Quantum Information via the project SciNet and the EU via the integrated project SIQS.

\appendix

\section*{Appendix}

\subsection*{Setting}

Our quantum repeater protocol works in the following way:
\begin{enumerate}
\item Bell pairs are generated in all elementary segments in a heralded way.
\item The Bell pairs are coupled to the resource states via Bell measurements. This implements the entanglement purification via hashing and the swap operations simultaneously.
\item The results from the Bell measurements are communicated to one of the repeater end stations via one-way classical communication.
\end{enumerate}

\begin{widetext}

\begin{figure}[ht]
\centering
\includegraphics[scale=0.7]{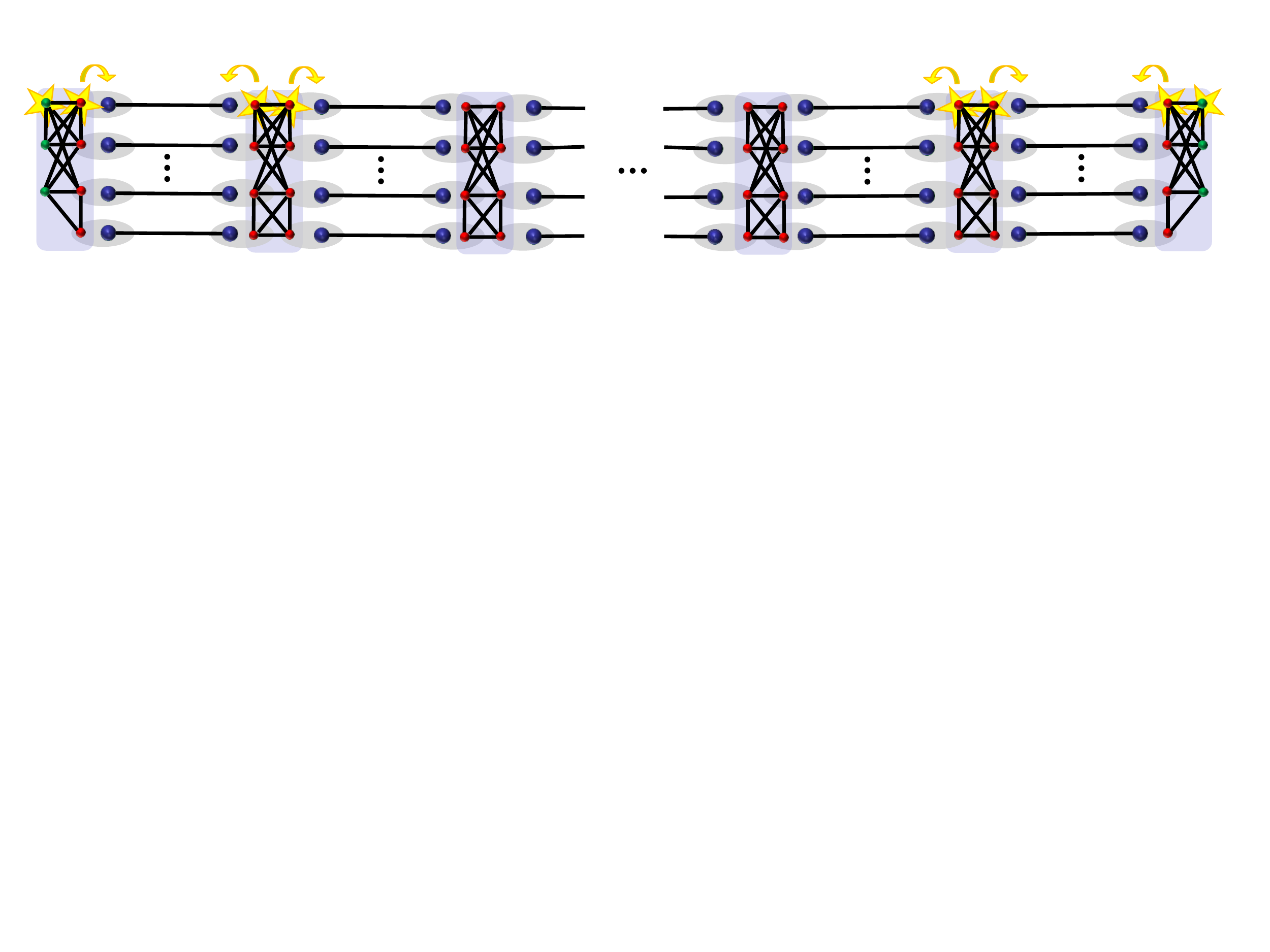}
\vspace{-10cm}
\caption{Illustration of a quantum repeater based on hashing. The channel is divided into $N$ elementary segments, where short-distance entangled pairs are generated over all segments, i.e., between all repeater stations, in parallel. Entanglement distillation via hashing and entanglement swapping are performed in a measurement-based way, by coupling the elementary pairs via Bell measurements to the locally stored resource state. In contrast to quantum repeaters based on recurrence protocols, no nested scheme is required. The purely QEC based version of our protocol consists in sending encoded information sequentially through the channel. We use standard graph state notation, but local unitaries are not shown. The initial Bell pairs are colored in blue, the qubits colored in red correspond to qubits of the resource states which are measured and the ones in green are the output of the protocol (the final Bell pairs). Please note that this is only an illustration, the real resource states contain at least order of one hundred qubits.}
\label{Figure_Setup_app}
\end{figure}

\end{widetext}

We would like to mention that it is possible to translate our scheme to a purely quantum error correction based quantum repeater \cite{Knill96}. This is due to the relation between quantum error correcting codes and entanglement purification protocols \cite{Be96}. The quantum error correcting code corresponding to the hashing protocol  encodes $m=cn$ logical qubits into $n$ physical qubits. The scaling of the local resources of such a scheme is similar, namely constant in the distance. A key difference is that two-way communication is not required (which one needs for the heralded generation of the initial Bell pairs, but only on a local scale). However, the time scale determining the ultimately achievable rate is in both cases given by the processing time $t_p$, see also the main text and the discussion below.

\subsection*{Construction of resource states}
In our quantum repeater architecture the processing of the Bell pairs is done in a measurement-based way. Instead of using a universal resource state \cite{Br2001,Ra2001}, we use optimized resource states, which only contain input and output qubits, which is possible since the underlying protocols (entanglement distillation via hashing and entanglement swapping) only involve Clifford gates and Pauli measurements. The resulting resource states are all graph states, up to local Clifford operations, and they can be determined in an efficient way \cite{Aaronson2004,Anders2005}. Please note that the resource states in Fig. \ref{Figure_Setup_app} are only for illustration, i.e., depicted are not the real graph states. The reason is that one needs resource states with at least around one hundred input qubits in order for hashing to work.

The construction of the resource states is completely analogous to \cite{Zw12,Zw13,Zw14,Zw14H}, for a review see \cite{Zw15}. We briefly discuss it below. Recall that the repeater stations combine two elementary tasks: entanglement distillation via hashing and entanglement swapping. 

Suppose there are $n$ short-distance Bell-pairs with entropy $S(W)$ between different repeater stations. According to the main text, the hashing protocol distills $m = n(1-S(W)-2 \delta)$ purified Bell-pairs from that ensemble. Hence we need to construct the resource state of the hashing protocol mapping $n$ noisy Bell-pairs to $m$ purified Bell-pairs which we denote by the map $H_{n \to m}$.

For that purpose we use the Jamio{\l}kowski isomorphism \cite{Jamiolkowski}, which establishes a one-to-one correspondence between completely positive maps and quantum states. In particular, for every quantum operation $\mathcal{O}$ there exists a, in general mixed, quantum state $\rho_{\mathcal{O}}$ which probabilistically implements the quantum operation $\mathcal{O}$ via Bell-measurements on the qubits which shall be processed and the input qubits of $\rho_{\mathcal{O}}$. This state, which we also refer to as resource state for the quantum operation $\mathcal{O}$, is given by 
\begin{align}
\rho_{\mathcal{O}} = (id \otimes \mathcal{O}) \left(\ket{\phi^+}\bra{\phi^+}\right)^{\otimes n}
\end{align}
where $n$ denotes the number of input qubits of $\mathcal{O}$. The state $\rho_{\mathcal{O}}$ is pure for maps involving Clifford gates and Pauli measurements and the measurement-based implementation of $\mathcal{O}$ is deterministic (since Pauli byproduct operators at the read-in can be propagated through the map $\mathcal{O}$). 

\begin{figure}[h!]
\begin{center}
\scalebox{7}{
\includegraphics{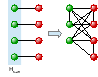}
}
%\flushleft
\caption[h!]{\label{fig:resource:hashing} Schematic construction of the resource state for the hashing protocol which maps $n$ noisy Bell-pairs to $m$ purified Bell-pairs. The green vertices correspond to the output qubits whereas the red vertices to the input qubits of the resource state. The light blue rectangle depicts the application of the hashing protocol $H_{n \to m}$ to one half of $n$ maximally entangled states. This resource state implements the hashing protocol $H_{n \to m}$ by performing a Bell-measurement between the noisy input Bell-pairs and the input qubits of the resource state.}
\end{center}
\end{figure}

Therefore we easily find, that the resource state of the hashing protocol $H_{n \to m}$ is given by 
\begin{align}
\ket{\psi_{H_{n \to m}}} = (id \otimes H_{n \to m}) \ket{\phi^+}^{\otimes n}
\end{align}
where $n$ denotes the number of input Bell-pairs of the hashing protocol, see Fig. \ref{fig:resource:hashing}.

\begin{figure}[h!]
\begin{center}
\scalebox{7}{
\includegraphics{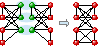}
}
%\flushleft
\caption[h!]{\label{fig:resource:repeater} The figure shows the construction of the resource state of the repeater stations, which is obtained by combining the resource states $\ket{\psi_{n \to m}}$ of the hashing protocol $H_{n \to m}$ via Bell-measurements (which implements the entanglement swapping operation).}
\end{center}
\end{figure}

From this we now construct the resource state for the repeater stations as follows: Recall that the repeater stations first run the hashing protocol $H_{n \to m}$ for their left and their right segment via two copies of the  resource state $\ket{\psi_{H_{n \to m}}}$ in a measurement-based way. Then they combine the  purified Bell-pairs, which correspond to the output qubits of the resource states $\ket{\psi_{H_{n \to m}}}$, of each segment via entanglement swapping, which amounts to a Bell-measurement and classically communicating the outcome. Therefore, we obtain the resource state of the repeater stations, which we denote by $\ket{\psi_{\mathrm{R}}}$, by performing a Bell-measurement between the output qubits of two resource states of the hashing protocol, see Fig. \ref{fig:resource:repeater}.
Since all operations involved belong to the Clifford group the resource state $\ket{\psi_{\mathrm{R}}}$ is of minimal size, i.e. it consists of $2n$ qubits. The resource state $\ket{\psi_{\mathrm{R}}}$ now implements entanglement distillation via hashing and entanglement swapping at repeater stations by coupling the short-distance Bell-pairs to the input qubits of $\ket{\psi_{\mathrm{R}}}$ via Bell-measurements, see Fig. \ref{Figure_Setup}. In addition, the repeater stations need to communicate the outcomes to one end station.

\subsection*{Noisy resource states}
We model imperfect resource states for measurement-based quantum computing in the following way: all qubits are affected by local depolarizing noise (LDN). LDN is defined by the map ${\cal{D}}$
\be
{\cal{D}}(p)\rho= p\rho + \frac{1-p}{4} \left(\rho+X\rho X + Y\rho Y + Z\rho Z \right).
\ee
The parameter $p \in [0, 1]$ quantifies the level of noise with $p=1$ corresponding to the noiseless case and $p=0$ to complete depolarization.. The density matrix of a noisy $n$-qubit resource state is then obtained by 
\be
\prod_{i=1}^{n} {\cal{D}}_i(p)\ket{G}\bra{G},
\ee
where $\ket{G}$ denotes the state vector of the graph state and the subscript indicates on which subsystem the respective map acts.

\subsection{Memory errors}
Here we argue that memory errors can be included in the noise acting on the resource states. We distinguish two cases. First, if one is interested in quantum key distribution, one can measure all qubits within the entire quantum repeater as soon as entanglement has been successfully generated. The time scale is given by $t_0$ (see main manuscript) and does not depend on the total distance. Thus the error arising from storing the resource states for $t_0$ can be included in the noisy resource state. Second, if one is interested in establishing Bell pairs, then the resource states at intermediate repeater stations still only need to be stored for the time scale $t_0$. The resource states at the outermost repeater stations (strictly speaking only the output qubits of these states) however need to be stored for the time scale $t_c$, determined by the classical communication, which scales linearly with the distance. Thus these qubits either have to be protected actively via quantum error correction or they need to have a coherence time larger than $t_c$. For earth based quantum repeaters, which we are mostly interested in, $t_c$ is at most $\mathcal{O}(10^{-1}) s$. There are already different experimental setups, where significantly larger coherence times have been achieved, see e.g. \cite{Roos2004,Haffner2005,Maurer1283,Saeedi830}.

\subsection{Rates}
Here we provide more information on the distribution times and rates. We distinguish between two different situations. First, a single-shot scenario where one creates long-distance Bell pairs once. Second, a continuos scenario, where one continuously establishes Bell pairs, e.g. for sending a number of qubits which is larger than what can be transferred in a single run of the repeater. The second scenario is more realistic and we will mostly focus on it. Here, it is possible to establish new elementary Bell pairs directly after the previous ones have been measured, and does not have to wait for the classical signal to arrive at the repeater end station. Thus, the classical communication time $t_c$ does not enter in the rate $R$. This is similar to QEC based quantum repeaters \cite{Knill96}, where the rate is also determined by the processing time of the logical qubits at each repeater station and not by the classical communication time. We always consider the rate per channel, that is, the ratio of the number of created Bell pairs per time unit and the number of parallel channels.

When using fewer parallel channels, say $[n\left(1/\eta + \epsilon\right)]/k$, $k$ repetitions of the pair creation process are required, resulting in time $k t_0 + t_c$ to generate $m=cn$ long-distance pairs. There is thus a direct tradeoff between the achievable rates (number of elementary pairs that are generated per second) and the number of parallel channels. Notice that, in contrast to a recurrence-based repeater, the overall rate is solely determined by the maximum of the waiting times over the elementary segments $t_0$ and the processing time $t_p$. Even with a single channel, we obtain the rate  $R_1=\tfrac{cn}{n(1/\eta + \epsilon)t_0 + t_p}$, which (in $n$) approaches a constant value $R_1 \approx \tfrac{c \eta}{t_0}$ determined only by the yield $c$ of the hashing protocol, and the average generation time of the elementary pair. Using $n_c \approx n/\eta$ parallel channels, the rate per channel is given by $R_{n_c}= \tfrac{c\eta}{t_0+t_p}$. Notice that $t_0$ can be made as small as the processing time $t_p$ by making the elementary segments short enough (it can not be made smaller since some entangling gate between a matter qubit and and a photon will be required). The normalized rates per channel $R_1$ and $R_{n_c}$ are identical, but the absolute rate is $n_c$ times higher when using $n_c$ parallel channels. Regarding local memory requirements, a $2n$ qubit resource state needs to be stored at each repeater station, in any case. However the number of extra qubits that need to be stored for the Bell pairs can be reduced to one, if one only uses a single channel. Then the local overhead is no longer $O = 4(1-S(W)-2\delta)^{-1}$, but rather $O = 2(1-S(W)-2\delta)^{-1}+1/n$, approaching $2(1-S(W))^{-1}$ in the asymptotic limit. Notice however that this comes at the expense of having to store the resource state for significantly longer time.

The multiplexing in the continuos scenario will not be possible for variants of the 1998 protocol with reduced memory requirements \cite{Duer1998,PhysRevA.72.052330,Childress2006}. Hence the rates for the continuos and the single-shot scenario will be the same.

In the single-shot scenario the time scales appearing above need to be changed to $t_0 + t_p + t_c$, which is dominated by $t_c$.

Finally we provide an example for the rate per channel, when the length of the elementary segments is $10$ km, the number of links is $N=1000$ (leading to a total distance of 10000 km), the speed of light is $c_{\rm{fiber}} = 2 \cdot 10^8 m/s$, $n=2000$, $\delta = n^{-1/4}$, $\epsilon= n^{-1/4}$, $\eta=2/3$, $t_p=1 \mu s$, the noise on the resource states is given by 1\% LDN, and the fidelity of the initial Bell pairs is $F=0.95$. In this case one obtains a rate per channel of $R_{example}\approx 3 \rm{kHz}$. Notice that one can make the rate substantially higher by going to shorter elementary segments and assuming more optimistic parameters. For example for a processing time of a nanosecond one could obtain a rate per channel of up to order of GHz.

\subsection*{Lower bound on the global fidelity}
The hashing protocol \cite{Be96} is an entanglement distillation protocol which operates on an asymptotically large ensemble of $n$ noisy Bell pairs. Information about the system is obtained from parity measurements on subsets of the ensemble. In the limit of infinitely many pairs ($n \to \infty$) the output of the protocol consists in $m = (1-S)n$ perfect Bell pairs, where $S$ denotes the (von Neumann) entropy of the initial Bell pairs. For more details see \cite{Be96}. In the following we are concerned with the failure probability of the hashing protocol which vanishes asymptotically but affects the fidelity of the output pairs in the case of a finite-size ensemble.

The total failure probability $p_{\rm{fail}}$ of the hashing protocol is bounded from above by the sum of two failure probabilities, $p_1$ and $p_2$. The first kind of failure, occurring with probability $p_1$, is that the classical bit-string corresponding to the randomly chosen subset of Bell pairs out of the initial ensemble falls outside the likely subspace ${\cal{L}}$, see fact (1) in III. B. 3 in \cite{Be96}. The second possibility for failure is given by the probability that two strings $x_r$ and $y_r$ remain distinct while having agreed on all $r$ subset parity measurements, see fact (3) in III. B. 3 in \cite{Be96} for more details. The probability of having more than one string surviving is then bounded by $2^{n[\text{S}(W) + \delta] - (n-m)}$ \cite{Be96}. \newline
Thus the failure probability $p_{\rm{fail}}$ provides a lower bound on the fidelity as a function of $n$ in the following way: if there is no error (which happens with probability $1-p_{\rm{fail}}$) the unmeasured pairs will all be in the $\ket{\phi^+}$ state. In any other case we assume that we obtain some unknown orthogonal state. Notice that the bound on the fidelity is a bound on the {\it global} fidelity, i.e., relative to a tensor-product state of $m$ perfect pairs. In a noisy measurement-based implementation one can  map noise on the input qubits of the resource state to the input Bell pairs, which effectively lowers their fidelity $F$, whereas the noise on the output qubits can be applied in the last step \cite {Zw13}. The fidelity that one obtains before the application of the noise on the output qubits is the global, private fidelity $F_{\rm{gp}}$. By applying the noise on the output qubits one can obtain bounds on the fidelities of the ensemble and the individual pairs. \newline
In a quantum repeater with many links one can get a bound on the final Bell pairs by considering the probability that the hashing protocol is successful in all links simultaneously. \newline

Now we turn to the proof of eq. (1) of the main text, and provide and provide more details on $\alpha$ and $\beta$. The global, private fidelity $F_{\rm{gp}}$ of the output pairs is bounded from below by $1-p_{\rm{fail}}$, where $p_{\rm{fail}}$ denotes the failure probability of the hashing protocol. This probability, in turn, is  bounded from above by the sum of the probability that the initial string falls outside the likely subspace, $p_1$, and the probability that two strings remain distinct while having agreed on all $r$ subset parity measurements, $p_2$. Here we explicitly estimate the bounds for $p_1$ and $p_2$ respectively, thereby proving eq. (2) of the main text. \newline
Before we provide these estimates, recall that the hashing protocol performs $n-m = n(S(W) + 2\delta)$ measurements to collect subset parity information of the ensemble. \newline
In order to derive a bound on the probability of falling outside the likely subspace $p_1$ one needs to consider so-called concentration inequalities, like e.g. Hoeffding's inequality \cite{Hoeffding} and Bennett's inequality \cite{BennettInequ}. Those inequalities have in common that they are mostly used to bound tail probabilities of independent random variables. Here we bound the probability $p_1$ via the Bennett inequality. Recall that the Bennett inequality \cite{BennettInequ} states that for $X_1,..,X_n$ independent and identically distributed (i.i.d.) random variables, where $|X_i| \leq a$ almost-surely and the expected value of $X_i$ is zero without loss of generality, that
\begin{align}
{\rm{Pr}} \left(\left|\sum^n_{i=0} X_i \right| > t \right) \leq 2 \exp\left( - \frac{n \sigma^2}{a^2} h\left(\frac{at}{n\sigma^2} \right) \right) \label{eq:bennett}
\end{align}
where $\sigma^2 = 1/n \sum_i \mathrm{Var} X_i$ and $h(u)=(1+u)\log(1+u)-u$. \newline
The random variables $X_i$ take the values  $X_i(k,l) := -\log_2 p_{kl} - S(W)$ in case of the hashing protocol where $p_{kl}$ denotes the probability of $\ket{B_{kl}} = (id \otimes \sigma_x^l \sigma_z^k) \ket{\phi^+}$ for $k,l \in \lbrace 0,1 \rbrace$ within the Bell-diagonal state $\rho = \sum^1_{k,l=0} p_{kl} |B_{kl}\rangle \langle B_{kl}|$ and $S(W) = - \sum^1_{k,l=0} p_{kl} \log_2 p_{kl}$. Observe that for states in Werner form, which we assume throughout this paper, we have $S(W) = -F \log_2 (F) - (1-F) \log_2 ((1-F)/3) =: S(F)$. Furthermore, because all random variables $X_i$ are independent and identical by our i.i.d. assumption, we easily find $\sigma^2 = 1/n \sum_i \mathrm{Var} X_i = \mathrm{Var} X =: V(F)$. We simplify $V(F)$ to
\begin{align*}
V(F) &= \mathrm{Var} X = \sum_{k,l} p_{kl} (-\log_2 p_{kl} - S(F))^2 \\
&= \sum_{k,l} p_{kl} (\log^2_2 p_{kl} + 2 S(F) \log_2 p_{kl} + S^2(F)) \\
&= \sum_{k,l} p_{kl} \log^2_2 p_{kl} + 2 S(F) p_{kl} \log_2 p_{kl} + p_{kl} S^2(F) \\
&= \sum_{k,l} p_{kl} \log^2_2 p_{kl} + 2 S(F)(-S(F)) + S^2(F) \\
&= F \log^2_2 F + (1-F) \log^2_2 ((1-F)/3) - S^2(F).
\end{align*}
We observe that $|X(k,l)| = |\log_2 p_{kl} + S(F)| \leq |\log_2((1-F)/3)| + S(F) =: a(F)$ because $|\log_2 ((1-F)/3)| > |\log_2 F|$ for $F > 0.8107$, which is the threshold for Werner states such that the hashing protocol works at all. \newline
We denote the left hand side of (\ref{eq:bennett}) by $p_1$. Setting $t=n\delta$ and inserting $a = a(F)$ and $\sigma^2 = V(F)$ we obtain
\begin{align*}
p_1 & \leq 2 \exp \left(\frac{-n V(F)}{a^2(F)} h \left(\frac{a(F)n \delta}{n V(F)} \right) \right) \\
&= 2 \exp \left \lbrace \frac{-n V(F)}{a^2(F)} \left[ \left(1 + \frac{a(F) \delta}{V(F)} \right) \log \left(1+\frac{a(F) \delta}{V(F)} \right) \right. \right. \\
& \left. \left. - \frac{a(F) \delta}{V(F)} \right] \right \rbrace \\
&= 2 \exp \left \lbrace \frac{-n}{a(F)} \left[ \left(\frac{V(F)}{a(F)} + \delta \right) \log \left(1+\frac{a(F) \delta}{V(F)} \right) - \delta \right] \right \rbrace.
\end{align*}
Defining $g(F) = \frac{V(F)}{a(F)}$ this becomes
\begin{align}
p_1 \leq 2 \exp \left \lbrace \frac{-n}{a(F)} \left[ \left(g(F) + \delta \right) \log \left(1+\frac{\delta}{g(F)} \right) - \delta \right] \right \rbrace. \label{eq:likely:outside}
\end{align}
Observe that (\ref{eq:likely:outside}) holds for all choices of $\delta$. \newline
The probability that two strings remain distinct after $n-m$ rounds while having agreed on all subset parities, $p_2$, is bounded by $p_2 \leq 2^{n[\text{S}(W) + \delta] - (n-m)}$ \cite{Be96}. Inserting $n-m=n(S(W) + 2 \delta)$ immediately yields
\begin{align}
p_2 & \leq 2^{-n \delta} \label{eq:toomany}.
\end{align}
Recalling that $F_{\rm{gp}} \geq 1 - p_{\rm{fail}}$ and $p_{\rm{fail}} \leq p_1 + p_2$ thus proves eq. (1) of the main text via the estimates (\ref{eq:likely:outside}) and (\ref{eq:toomany}) for $p_1$ and $p_2$ respectively, i.e.,
\begin{align*}
F_{\rm{gp}} \geq 1  - 2 e^{\left \lbrace \tfrac{-n}{a(F)} \left[ \left(g(F) + \delta \right) \log \left(1+\tfrac{\delta}{g(F)} \right) - \delta \right] \right \rbrace } 
 - 2^{-n \delta}
\end{align*}
as to be proven.

We would like to mention that one can drop the assumption that the input Bell pairs are i.i.d. using similar methods as in \cite{Zwerger17}.

\subsection{Additional data on reachable fidelity and yield}

\subsubsection{$n \to m$ Hashing}
\begin{figure*}[ht!]
 \centering
 \subfloat[\centering]{\includegraphics[width=0.48\columnwidth]{fidelity15100-eps-converted-to.pdf}}
 \hfill
 \subfloat[\centering]{\includegraphics[width=0.48\columnwidth]{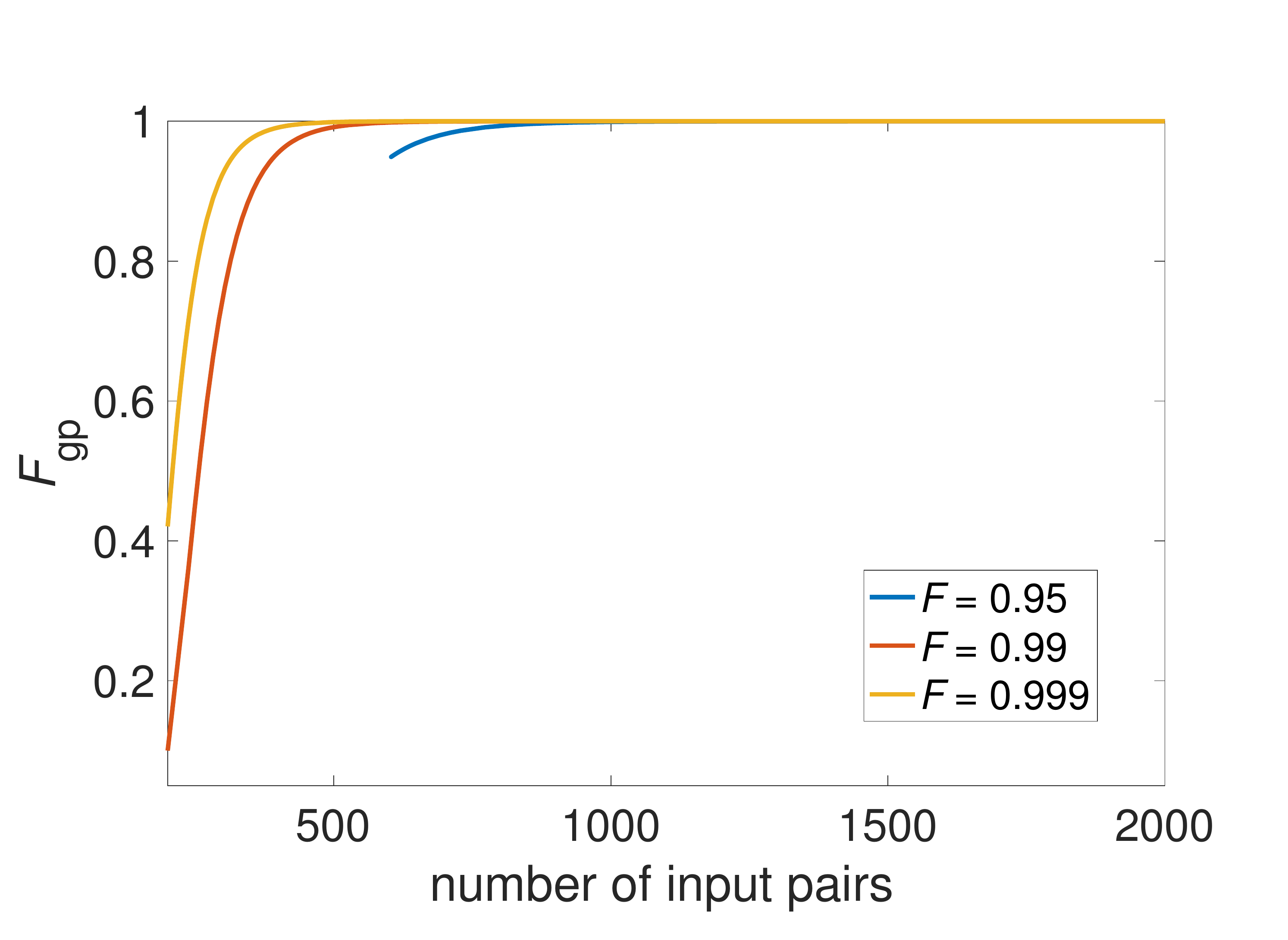} }
 \hfill
 \subfloat[\centering]{\includegraphics[width=0.48\columnwidth]{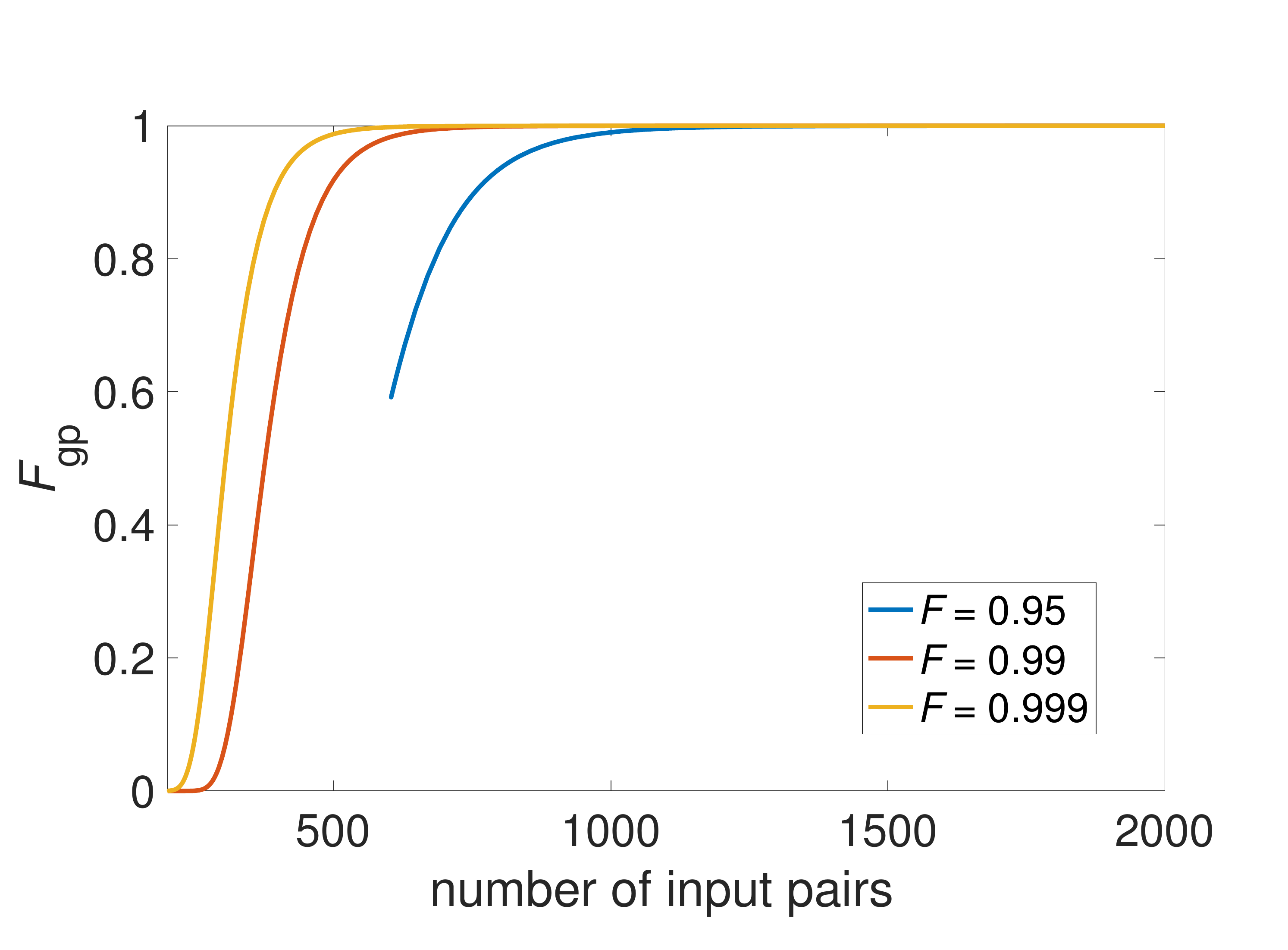} }
 \hfill
 \subfloat[\centering]{\includegraphics[width=0.48\columnwidth]{y15-eps-converted-to.pdf}}

 \vspace{-3mm}

 \centering
 \subfloat[\centering]{\includegraphics[width=0.48\columnwidth]{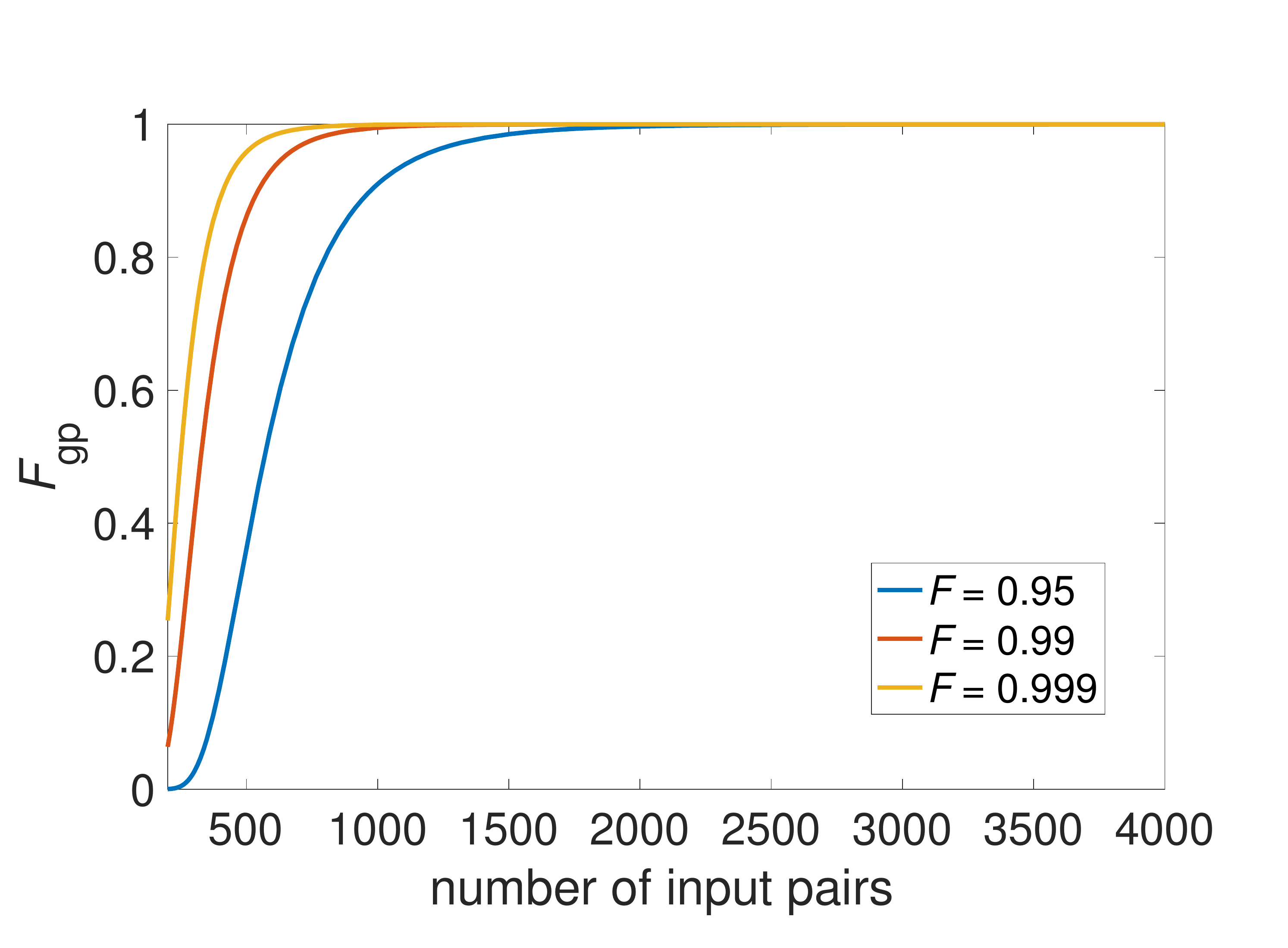}}
 \hfill
 \subfloat[\centering]{\includegraphics[width=0.48\columnwidth]{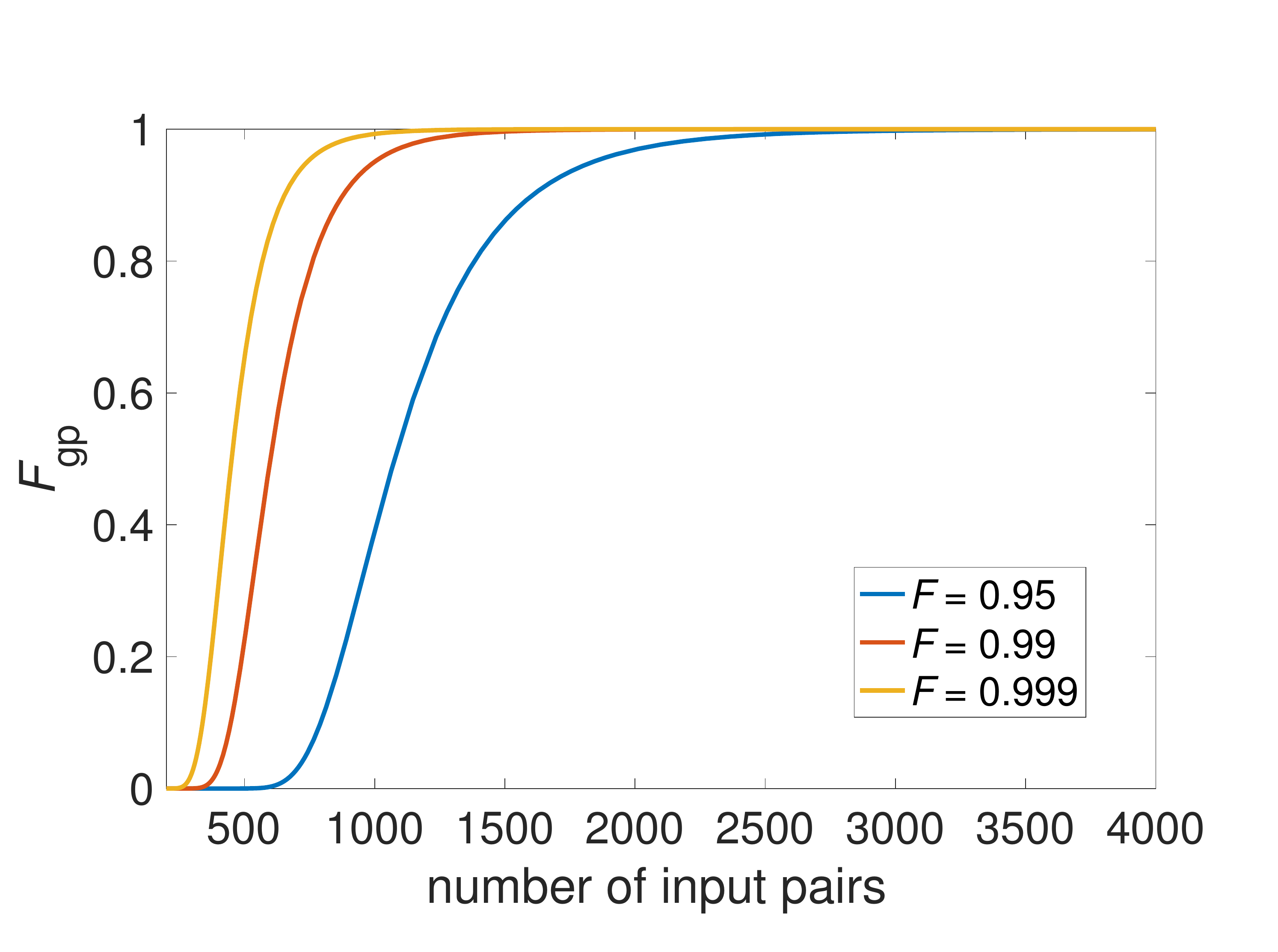} }
 \hfill
 \subfloat[\centering]{\includegraphics[width=0.48\columnwidth]{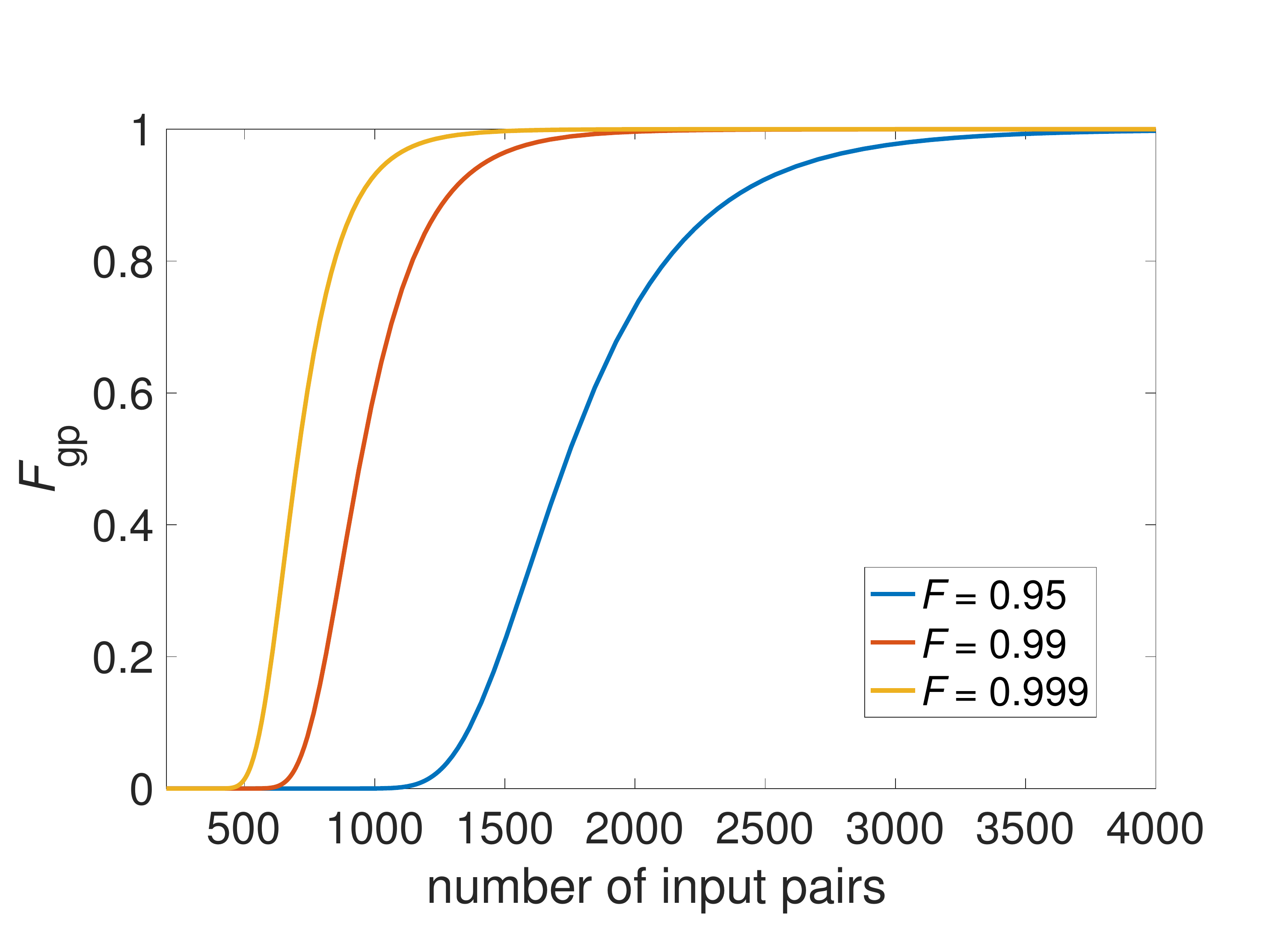} }
 \hfill
 \subfloat[\centering]{\includegraphics[width=0.48\columnwidth]{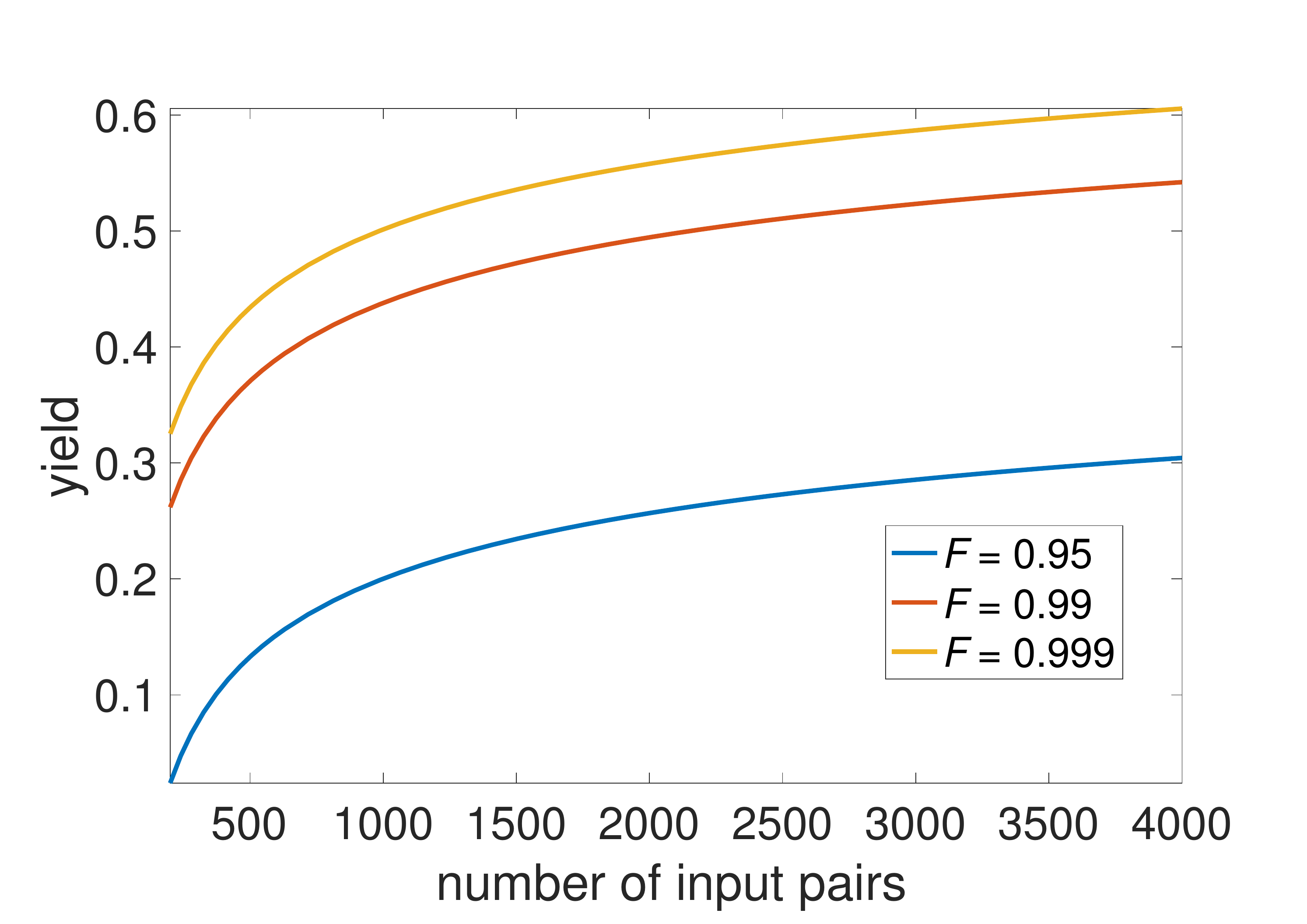}}

 \vspace{-3mm}

\centering
 \subfloat[\centering]{\includegraphics[width=0.48\columnwidth]{fidelity13100-eps-converted-to.pdf}}
 \hfill
 \subfloat[\centering]{\includegraphics[width=0.48\columnwidth]{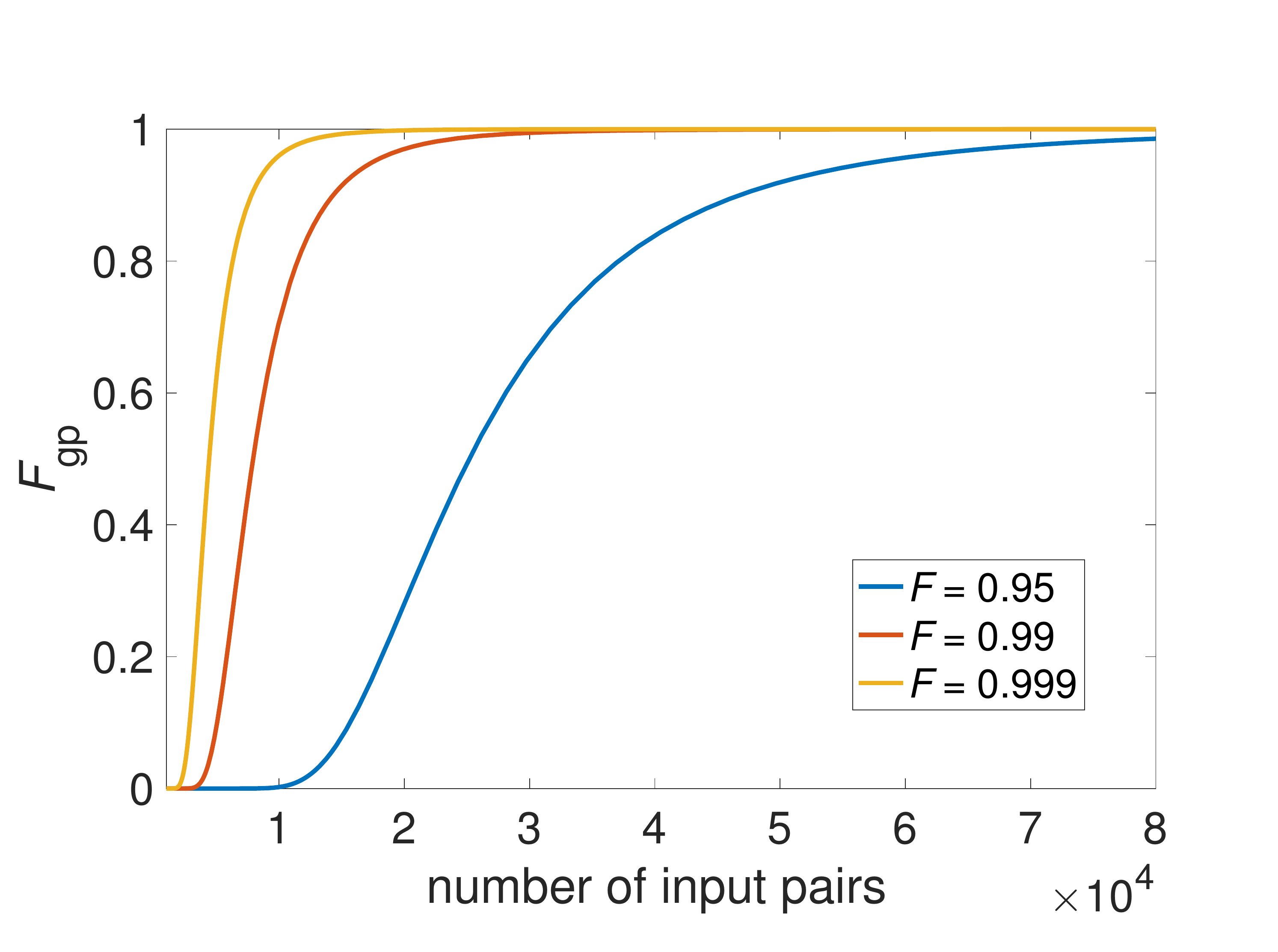} }
 \hfill
 \subfloat[\centering]{\includegraphics[width=0.48\columnwidth]{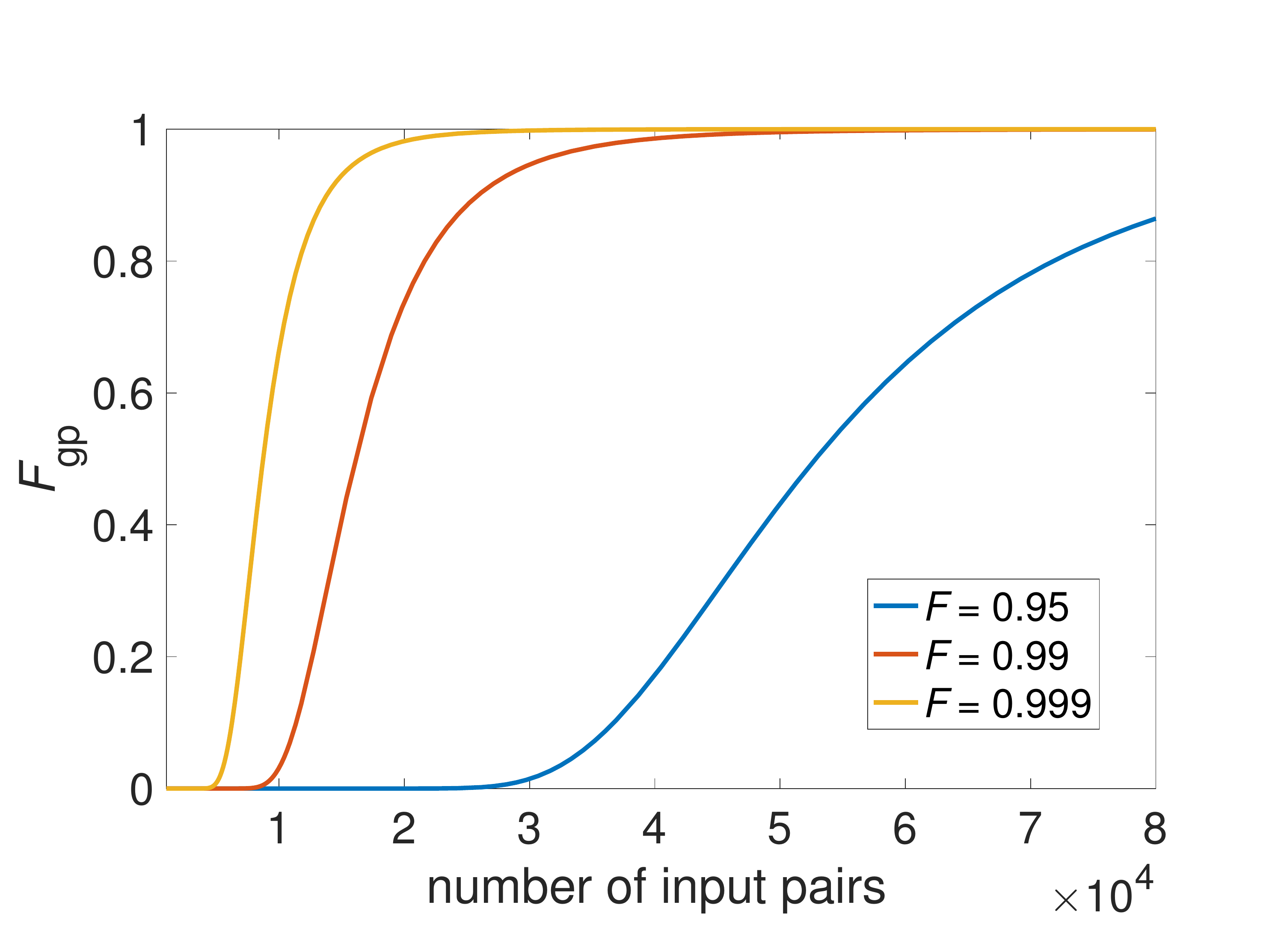} }
 \hfill
 \subfloat[\centering]{\includegraphics[width=0.48\columnwidth]{y13-eps-converted-to.pdf}}

 \caption{Plot of the private, global fidelity and yield as a function of the number of initial pairs for $\delta=n^{-1/5}$ (first line, (a,b,c,d)), $\delta=n^{-1/4}$ (second line, (e,f,g,h)) and $\delta=n^{-1/3}$ (third line, (i,j,k,l)). The number of repeater links is $N=100$ (first column, (a,e,i)), $N=1000$ (second column, (b,f,j)), and $N=10000$ (third column, (c,g,k))). The fourth column (d,h,l) shows the yield. We assume local depolarizing noise of $1\%$ per qubit for noisy resource states, and $F$ denotes the fidelity of the initial Bell pairs.}
 \label{FigF_Yield_allN}
 \vspace{-3mm}
\end{figure*}

In this section we provide additional numerical results on the reachable fidelity and yield for a hashing-based repeater with finite number of copies $n$. We consider different numbers of links, namely $N=100, N=1000, N=10000$ and different values of $\delta$. As discussed in the main text, the choice of $\delta$ influences the reachable fidelity and the yield, where there is a tradeoff between these two quantities. Notice that the yield is independent of the communication distance and hence the number of links $N$. The results are shown in Fig. \ref{FigF_Yield_allN}. Notice that the number of input pairs $n$ required to reach a certain value of $F_{\rm{pg}}$ grows only logarithmically with the number of links $N$ (see eq. (1) in the main text). However, the overhead per transmitted qubit approaches a constant, since a larger number of input pairs leads to a larger number of output pairs.

\subsubsection*{$n \to 1$ Hashing}

In this section we provide results for the (noiseless) $n \to 1$ hashing protocol.  \newline

We now have a closer look at the hashing protocol and the fidelity of the output pairs if the number of output pairs is varied. One can give a lower bound on the global fidelity of the output Bell pairs of hashing, assuming that the initial Bell pairs are in Werner form \footnote{one can always enforce this via depolarization \cite{Be96}}, with
\begin{align}
F_{\rm{gp}} \geq 1  - 2 e^{\left \lbrace \tfrac{-n}{a(F)} \left[ \left(g(F) + \delta \right) \log \left(1+\tfrac{\delta}{g(F)} \right) - \delta \right] \right \rbrace }
 - 2^{-n \delta} \label{eq:nto1}
\end{align}
where $F$ denotes the initial fidelity, $a(F) = |\log_2((1-F)/3)| + S(W)$ and $g(F) = [F \log^2_2 F + (1-F) \log^2_2 ((1-F)/3) - S^2(W)]/a(F)$. %For a detailed proof of (\ref{eq:nto1}) we refer to \cite{Sup}. 
We emphasize that the noise acting on the input qubits of the resource state is incorporated in $F$ and the noise of strength $p$ acting on the output qubits still needs to be applied.\newline

Recall that the hashing protocol performs $n-m = n(S(W) + 2\delta)$ measurements where $m$ is the number of output pairs and $S(W)$ denotes the von-Neumann entropy of the initial ensemble. Since $m=1$ for a single output pair, we find $\delta = 1/2((n-1)/n - S(W))$. Consequently this choice of $\delta$ provides a lower bound on the fidelity $F'$ (which equals $F_{\rm{gp}}$ in the noiseless case) of the output pair via (\ref{eq:nto1}).
\begin{figure}
 \centering
 \subfloat[\centering]{\includegraphics[width=0.48\columnwidth]{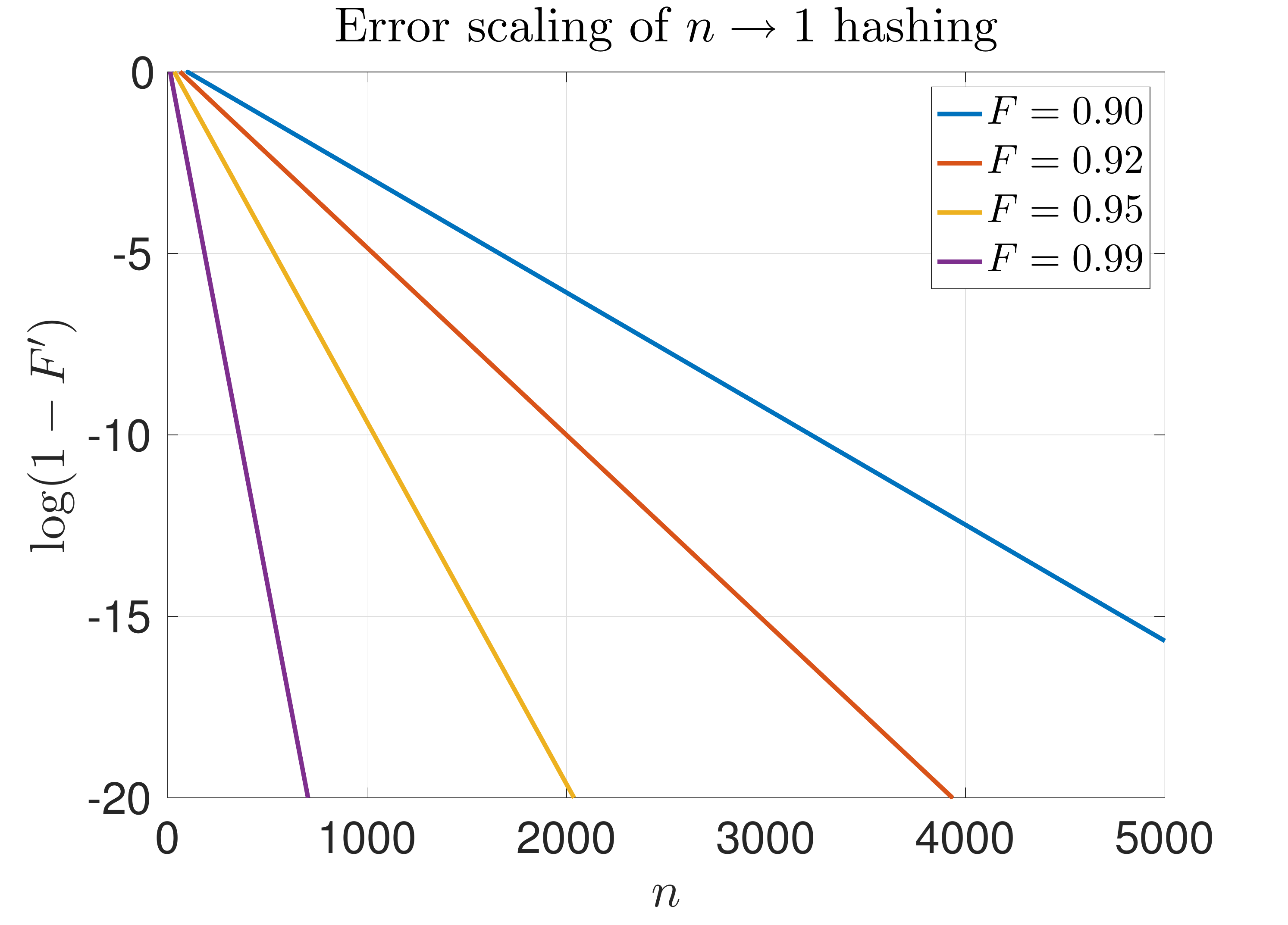} \label{fig:nto1:scale}}
 \hfill
\subfloat[\centering]{\includegraphics[width=0.48\columnwidth]{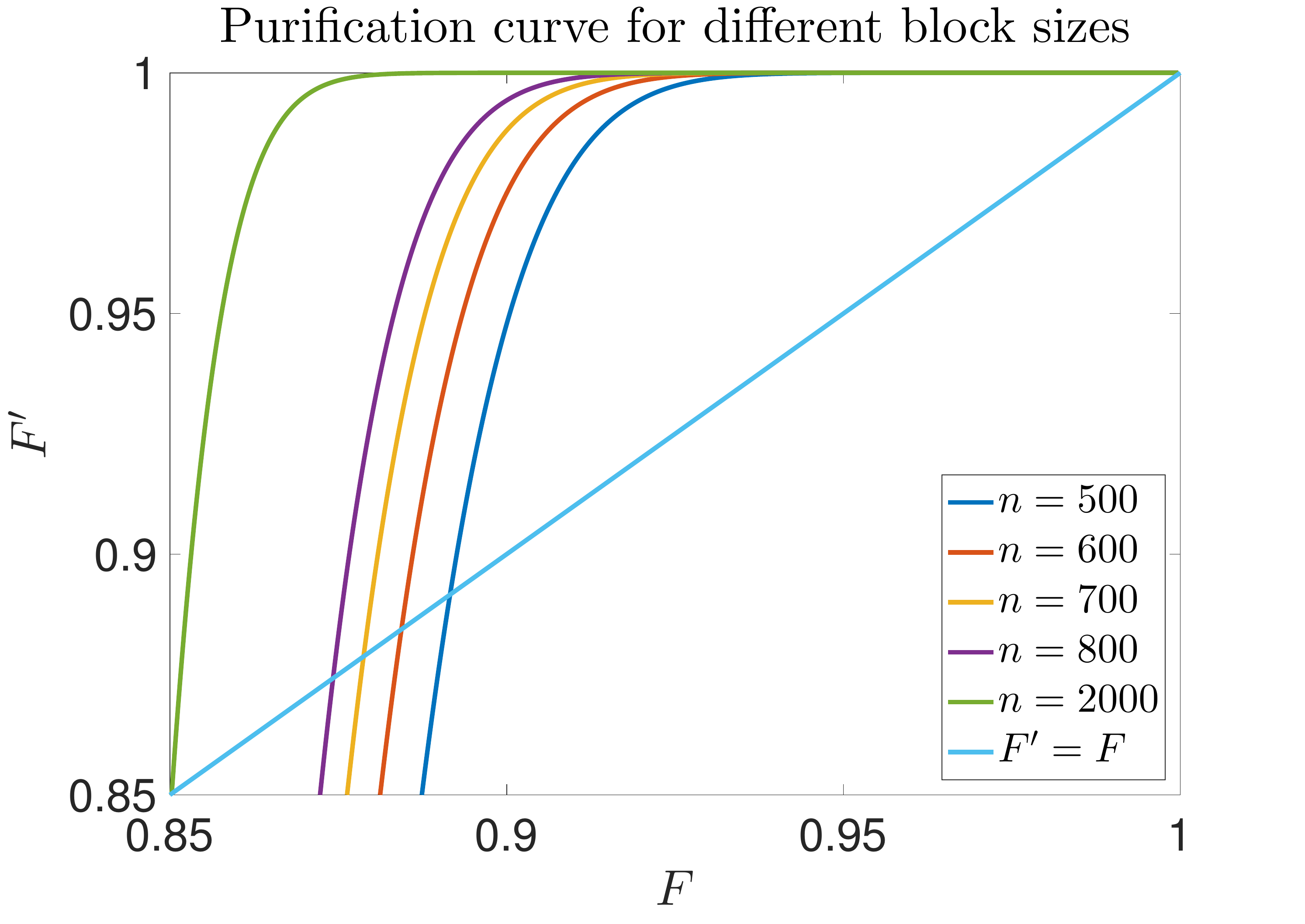} \label{fig:nto1:pur}}
 \caption{The figures summarize the results for the $n \to 1$ hashing protocol for a single link: Fig. \ref{fig:nto1:scale} plots the deviation in the output fidelity depending on the number of initial pairs for different initial fidelities. In Fig. \ref{fig:nto1:pur} the input fidelity versus the output fidelity for varying numbers of initial pairs is shown.}
 \label{fig:nto1}
 \vspace{-3mm}
\end{figure}

The results for the (noiseless) $n \to 1$ hashing protocol are summarized as follows: As the fidelity of the initial pairs tends to $1$, fewer initial pairs are necessary to achieve purification. For example, for initial fidelity $F=0.85$,  hashing requires at least $2027$ pairs to guarantee purification, whereas, for $F=0.95$,  $164$ pairs suffice. Fig. \ref{fig:nto1:scale} shows the exponential scaling governed by the number of initial pairs $n$ (see Eq. \ref{eq:nto1}). The purification curve, Fig. \ref{fig:nto1:pur}, shows that for a fixed input fidelity, larger ensembles lead to higher output fidelities, which is intuitive.

In order to reduce memory requirements, one may also consider a concatenated implementation of the $n \to 1$  hashing protocol \cite{Be96} using fixed input blocks of size $n$ (and a twirl step, to ensure the Werner form). In contrast to standard recurrence protocols \cite{Be96a,De96} this deterministically yields a high fidelity Bell pair.
Finally, we estimate the overhead for connecting $N$ segments using $n \to 1$ hashing. For that purpose we set $\delta = 1/2((n-1)/n - S)$ in (\ref{eq:nto1}) and observe that $F' \geq 1 - \alpha \exp(-\beta n)$ from Fig. \ref{fig:nto1:scale}. Therefore, as in (\ref{eq:nlinks}), the fidelity $F'$ after connecting $N$ segments satisfies $F' \geq 1 - N \alpha \exp(-\beta n)$. We can approximate the number of initial pairs necessary to connect $N$ segments with fidelity $F'$ using $n \approx  \beta^{-1} \log (\alpha N / (1-F'))$ where $\alpha$ and $\beta$ depend on the initial fidelity $F$. That is, there is a logarithmic overhead in required local resources per final pair with the distance. For example, to guarantee purification, i.e. $F' \geq F$, for $F=0.99$ and $N=100$ links, at least $n \approx 151$ initial pairs are necessary. \newline
We contrast this to $n \to m$ hashing, where the situation is different since with an increasing number of initial pairs we obtain an increasing number of output pairs, resulting in a {\it constant} overhead per transmitted qubit. \newline

\begin{figure}[htb]
 \includegraphics[scale=0.3]{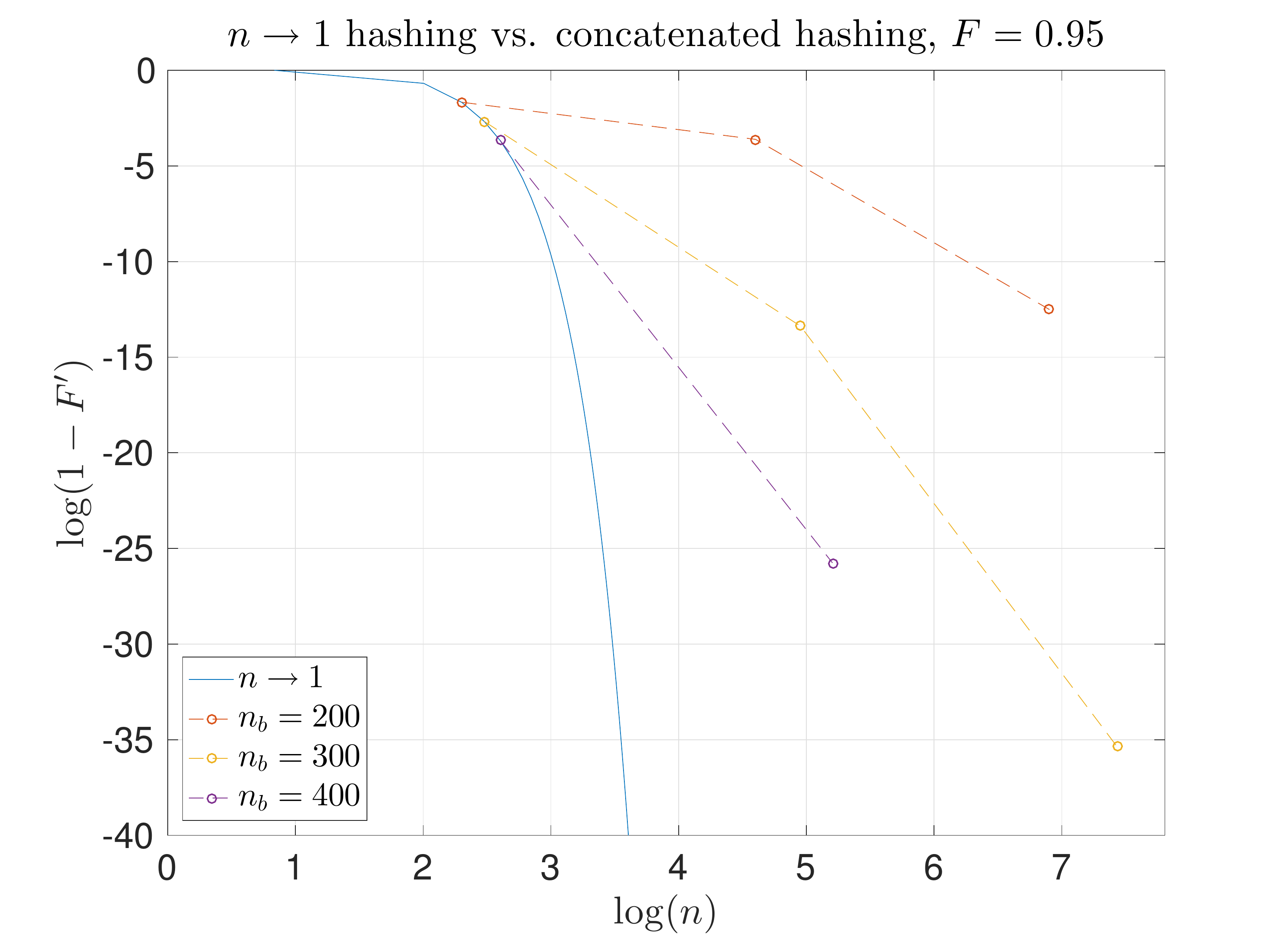}
 \caption{This plot compares the $n \to 1$ hashing protocol to a concatenated implementation of the hashing protocol with different blocks sizes. The achieved fidelity of the output pair (logarithmic scale) is plotted against the total number of resources, i.e., initial pairs. The initial fidelity of the Werner states is $F = 0.95$ and the block sizes shown are $n_b \in \lbrace 200, 300, 400 \rbrace$.}
 \label{fig:app:nto1:concvsnto1}
\end{figure}

\begin{figure}[htb]
 \includegraphics[scale=0.3]{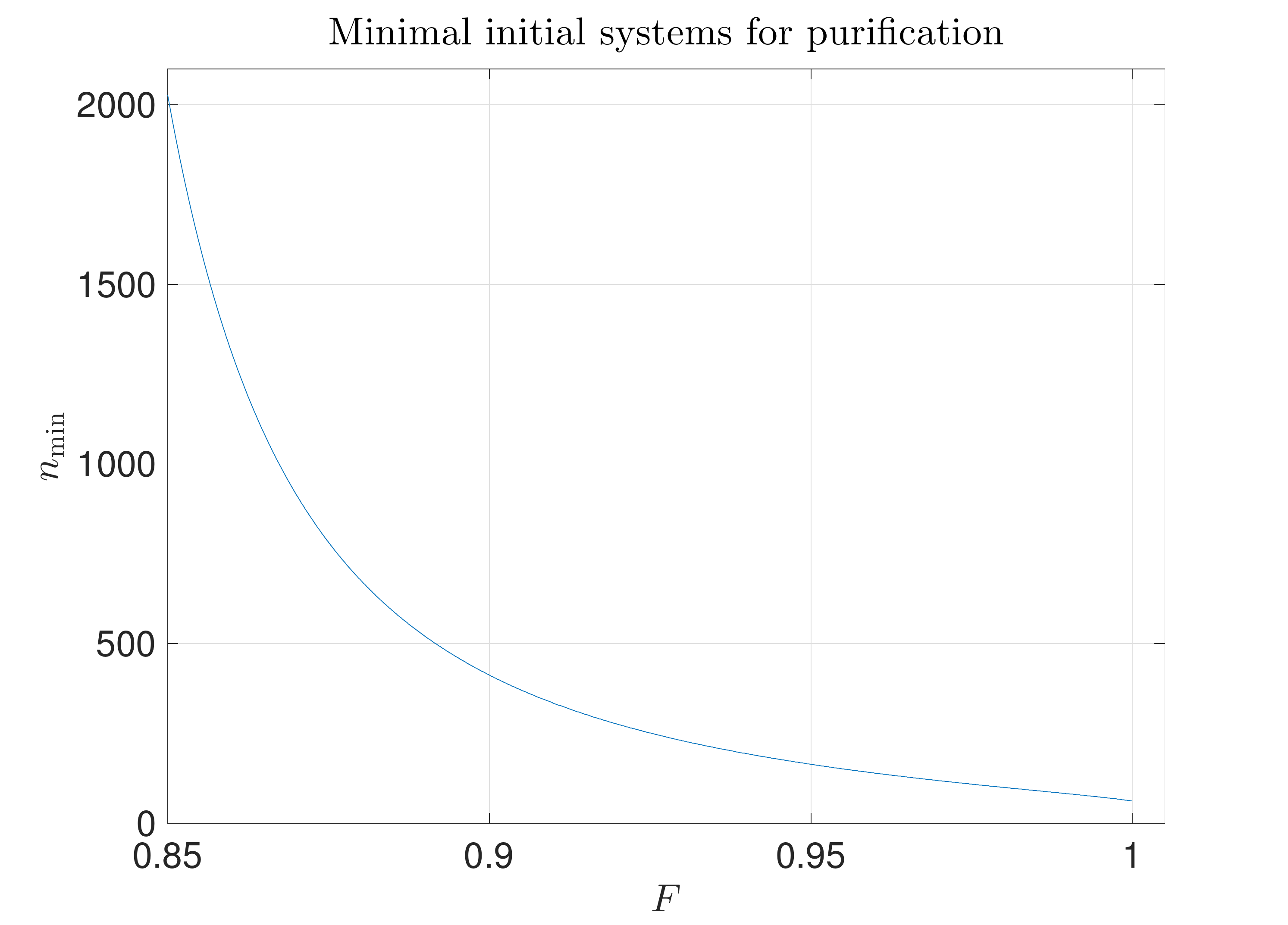}
 \caption{Plot of the minimum number of initial pairs in Werner form such that purification is feasible as a function of the initial fidelity.}
 \label{fig:app:nto1:nmin}
\end{figure}

\begin{figure}[htb]
 \includegraphics[scale=0.3]{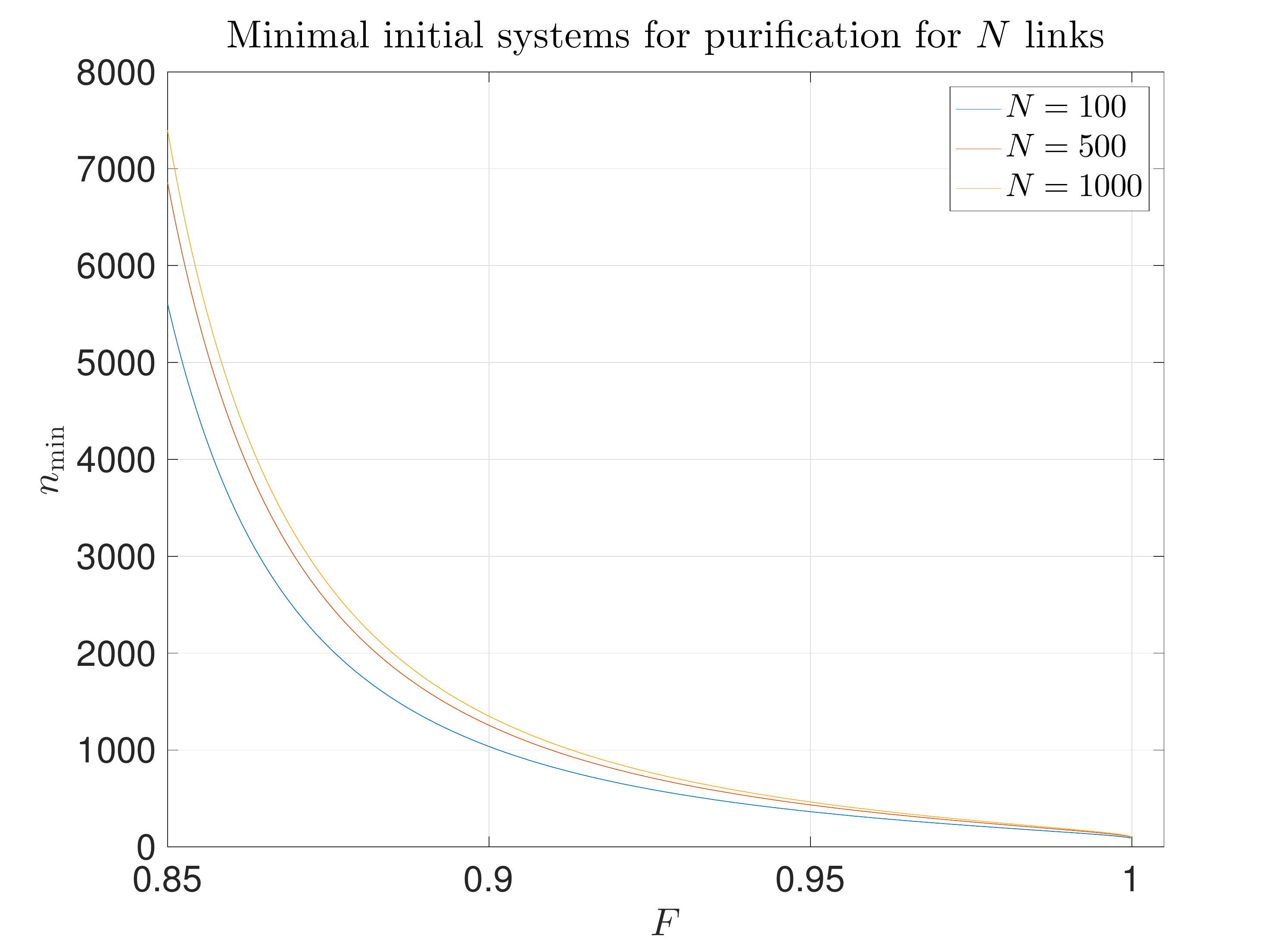}
 \caption{The figure shows the minimum number of initial pairs in Werner form such that purification is feasible after connecting $N$ links, as a function of  on the initial fidelity.}
 \label{fig:app:nto1:nminlinks}
\end{figure}

\begin{figure}[htb]
 \includegraphics[scale=0.3]{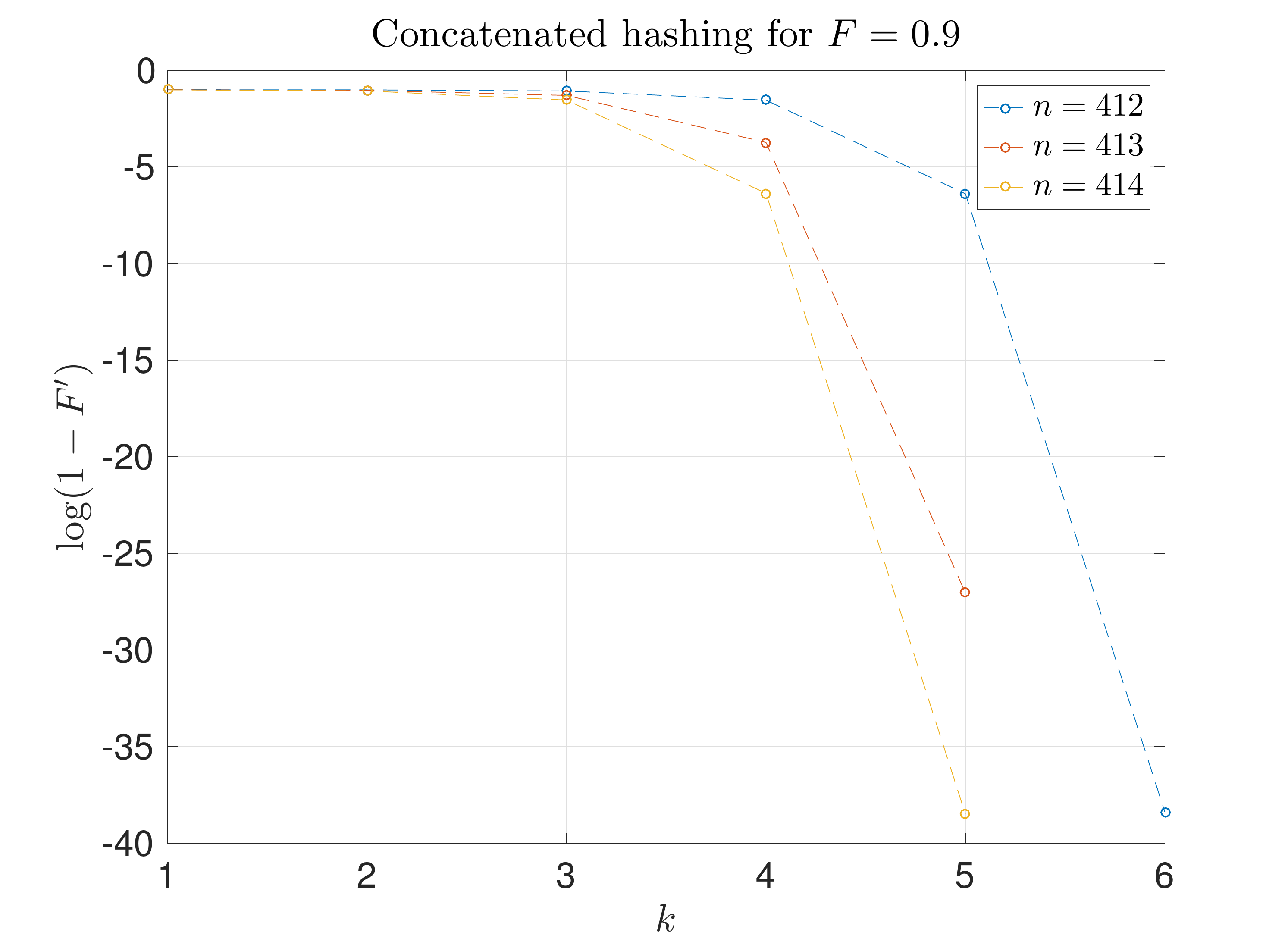}
 \caption{Plot of the rate of convergence of the concatenated $n \to 1$ hashing protocol for initial fidelity $F=0.9$ and different block sizes.}
 \label{fig:app:nto1:conc}
\end{figure}

Fig. \ref{fig:app:nto1:concvsnto1} compares direct $n \to 1$ hashing with concatenated implementations thereof. For that purpose we append a twirl towards Werner form after each concatenation level. On the one hand, the concatenated implementation with fixed block size has the advantage that less qubits need to be stored temporarily. In addition, as mentioned in the main text, the classical side-processing problem to evaluate the hash function is apparently hard \cite{Lo09}, which might become relevant if block sizes are too big. A concatenated application of $n \to 1$ hashing with moderate block sizes $n$ allows one to circumvent this problem.
On the other hand, we immediately infer that the rate of convergence for such a concatenated protocol is significantly worse compared to a direct $n \to 1$ approach where hashing is performed on a big ensemble. As expected, larger block sizes lead to a higher output fidelity.
In Fig. \ref{fig:app:nto1:nmin} we provide a plot where the minimum number of initial pairs $n_{\rm{min}}$ such that $F'(n_{\rm{min}}) \geq F$ is shown, i.e., distillation is guaranteed by the $n \to 1$ hashing. The plot suggests an exponential relationship between initial fidelity and the minimal number of  required initial pairs. For example, from the plot we observe that for initial fidelity $F \approx 0.9$ approximately $n_{\rm{min}} \approx 410$ initial pairs are required for distillation. As intuitively expected, the higher the initial fidelity the less initial pairs are necessary for purification. \newline
The situation turns out to be similar if we connect $N$ segments via entanglement swapping at the intermediate quantum repeater stations, see Fig. \ref{fig:app:nto1:nminlinks}.
Finally we provide a plot of the concatenated $n \to 1$ hashing for different block sizes in Fig. \ref{fig:app:nto1:conc}. Here, the initial fidelity is $F=0.9$ and the plot shows the rate of convergence for the block sizes $n \in \lbrace 412, 413, 414 \rbrace$. From that we observe that already a small increase in the block size, e.g. a single qubit, leads to massive improvement in the resulting output fidelity.

\subsection{Comparison of approaches}
The error thresholds for measurement-based entanglement purification and quantum error correction \cite{Zw12,Zw13,Zw14,Zw14H} - i.e. the tasks required for quantum repeaters - are around a few percent for a direct implementation of the Deutsch \textit{et al}. protocol, around ten percent for quantum error correction and the hashing protocol and around 20 percent for an optimized implementation of the Deutsch \textit{et al}. protocol. The tolerable noise is thus comparable, since optimized measurement-based implementations can all tolerate of the order of ten percent local depolarizing noise.

The main, qualitative difference between our scheme and existing ones is the superior scaling of the local resources. For an overview over key features of different quantum repeater architectures see the table in the main text.

Next, we compare the achievable rates for a measurement-based implementation of the 1998 protocol \cite{Br98} and our new scheme. The noise on the resource state is in both cases $1\%$, the fidelity of the initial Bell pairs and the number of elementary links is varied. 

We assume that $M$ Bell pairs are initially created in each segment and subsequently processed. We neglect the time for creating these pairs, because it is small compared to the classical communication over the entire channel. This follows from the fact that this time will be of the order of the classical communication time for an elementary segment for short segments with low absorption probability and for light-matter interfaces with high success probability. The total classical communication time however is at least $2^{N}$ times larger, where $2^{N}$ is the number of segments. We consider the cases $N=7$ to $N=13$. We also neglect the local processing time, which for our new protocol is given by the time to perform a Bell measurement, and for 1998 protocol is given by the time to perform a sequence of Bell measurements (to couple the output qubits of one round of entanglement distillation to the resource state for the next round). Thus the comparison is in favor of the 1998 protocol.

Then we compute the time it takes to create Bell pairs over the entire channel, both for the 1998 protocol and our new protocol. The rate is determined by the number of established Bell pairs divided by the time to create them and the number of initial pairs in each segment ($M$), and is given in units of $t_{\rm{segment}}^{-1}$. Here, $t_{\rm{segment}}$ is the time which classical communication over a single segment takes. Our results are summarized in tables \ref{tablerate1} and \ref{tablerate2}. On the one hand, the rates of our new quantum repeater scheme are up to nine orders of magnitude higher. On the other hand, the fidelity of the established Bell pairs is also higher.

For QEC based quantum repeaters \cite{Knill96} the rate is limited by the processing time $t_p$ (in the continuous scenario), similarly to our approach. In order to achieve achieve intercontinental distances with around $10000$ segments, a logical qubits needs to be encoded into several hundred physical qubits \cite{Muralidharan2014}. In our scheme only around two elementary Bell pairs in each segment are needed on average to establish a final, long-range Bell pair. Thus we anticipate that the rates for our approach are two to three orders of magnitude higher than the ones for QEC based quantum repeaters. However, we would like to mention that the error models and parameters in \cite{Muralidharan2014} are not directly comparable to the ones we use here. We leave a detailed comparison to future research.

Concerning the implementation of the 1998 protocol we have minimized the number of local resources $M$ for each choice of $N$ and initial fidelity $F$ as a function of the working fidelity (see Fig. \ref{W095} and Fig. \ref{W099}). The working fidelity is the fidelity up to which one distills the Bell pairs before swapping them, for more details see \cite{Br98}.

For the hashing protocol used in our scheme we have chosen $\delta=n^{-1/4}$.

We would like to mention that rates are calculated for a single shot scenario (see above). In the more realistic continuous scenario, the differences will be even more extreme. This is because in our scheme one can already start to establish new elementary Bell pairs after the ones from the previous round are processed and does not need to wait for global classical communication, in contrast to a repeater with two-way classical communication \cite{Br98}.

\begin{widetext}

\begin{table}
\caption{Comparison of rates and fidelities of output Bell pairs $F_{\rm{out}}$ for the 1998 quantum repeater protocol and our new quantum repeater protocol. We assume an input fidelity $F=0.95$. The number in brackets indicates the protocol. The rate is given in units of $t_{\rm{segment}}^{-1}$.}
\vspace{0.5cm}
\centering
\begin{tabular}{| c | c | c | c | c | c | c | c |}
\hline
$\#$ links & $2^7$ & $2^8$ & $2^9$ & $2^{10}$ & $2^{11}$ & $2^{12}$ & $2^{13}$ \\
\hline
rate (1998) & $7.186\cdot10^{-8}$ & $6.502\cdot10^{-9}$ & $6.400\cdot10^{-10}$ & $7.467\cdot10^{-11}$ & $7.370\cdot10^{-12}$ & $6.625\cdot10^{-13}$ &  $7.776\cdot10^{-14}$ \\
\hline
rate (2017) & $3.141\cdot10^{-3}$ & $1.778\cdot10^{-3}$ & $9.547\cdot10^{-4}$ & $4.973\cdot10^{-4}$ & $2.562\cdot10^{-4}$ & $1.308\cdot10^{-4}$ & $6.613\cdot10^{-5}$ \\
\hline
$F_{\rm{out}}$ (1998) & $0.8956$ & $0.9033$ & $0.8844$ & $0.8956$ & $0.9033$ & $0.9100$ & $0.8956$\\
\hline
$F_{\rm{out}}$ (2017) & $0.9851$ & $0.9851$ & $0.9851$ & $0.9851$ & $0.9851$ & $0.9851$ & $0.9851$ \\
\hline
\end{tabular}
\label{tablerate1}
\end{table}

\begin{table}
\caption{Comparison of rates and fidelities of output Bell pairs $F_{\rm{out}}$ for the 1998 quantum repeater protocol and our new quantum repeater protocol. We assume an input fidelity $F=0.99$. The number in brackets indicates the protocol. The rate is given in units of $t_{\rm{segment}}^{-1}$.}
\vspace{0.5cm}
\centering
\begin{tabular}{| c | c | c | c | c | c | c | c |}
\hline
$\#$ links & $2^7$ & $2^8$ & $2^9$ & $2^{10}$ & $2^{11}$ & $2^{12}$ & $2^{13}$ \\
\hline
rate (1998) &  $8.162\cdot10^{-7}$  & $7.264\cdot10^{-8}$ & $6.596\cdot10^{-9}$ & $6.551\cdot10^{-10}$ & $7.213\cdot10^{-11}$ & $7.331\cdot10^{-12}$ & $6.614\cdot10^{-13}$ \\
\hline
rate (2017) & $4.003\cdot10^{-3}$ & $2.382\cdot10^{-3}$ & $1.316\cdot10^{-3}$ & $6.967\cdot10^{-4}$ & $3.607\cdot10^{-4}$ & $1.847\cdot10^{-4}$ & $9.394\cdot10^{-5}$ \\
\hline
$F_{\rm{out}}$ (1998) & $0.8656$ & $0.8900$ & $0.9011$ & $0.8656$ & $0.8900$ & $0.9011$ & $0.9089$ \\
\hline
$F_{\rm{out}}$ (2017) & $0.9851$ & $0.9851$  & $0.9851$ & $0.9851$ & $0.9851$ & $0.9851$ &  $0.9851$ \\
\hline
\end{tabular}
\label{tablerate2}
\end{table}

\begin{figure}[htb]
\includegraphics[scale=0.45]{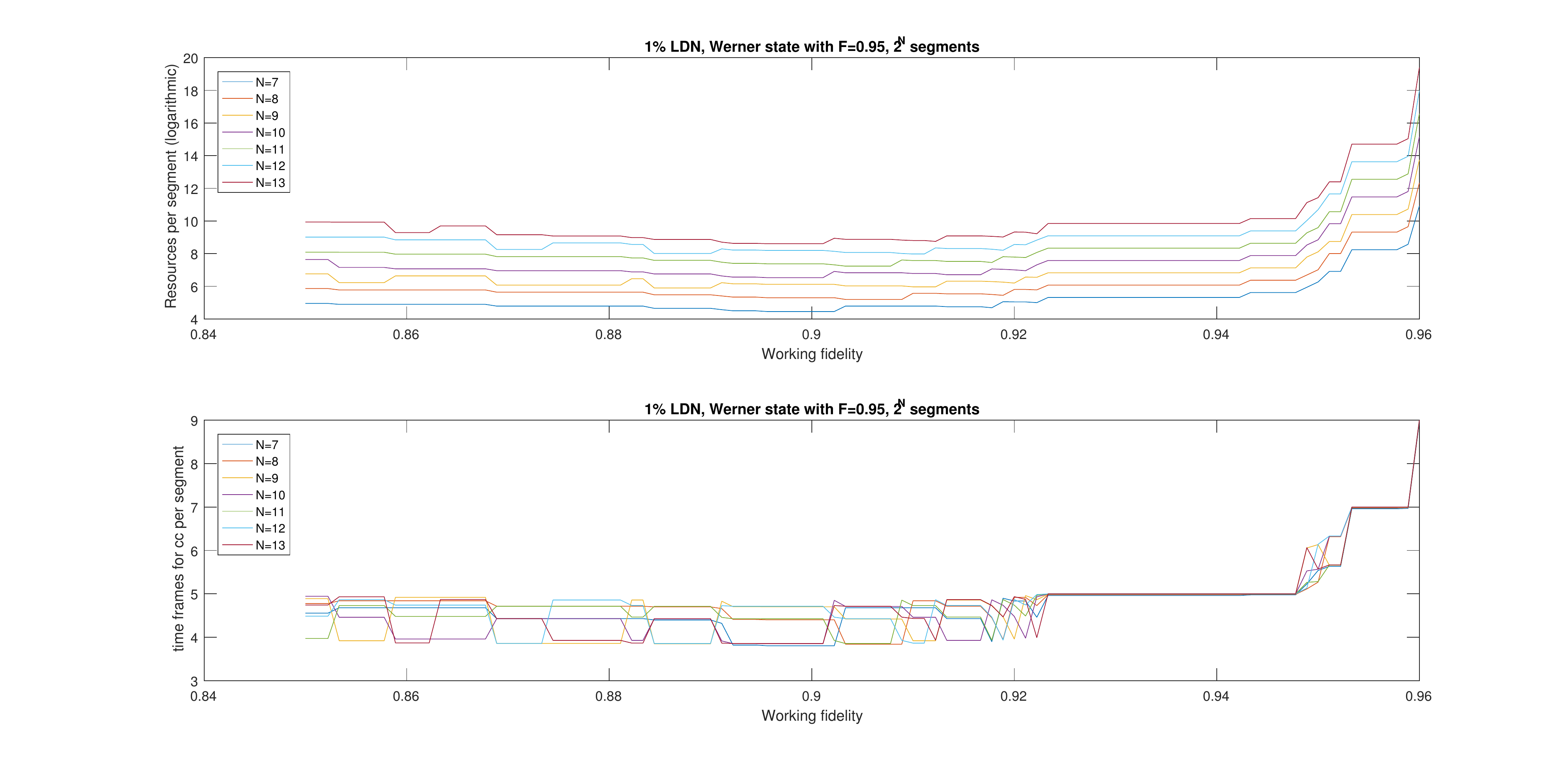}
\caption{(top) Plot of the number of elementary Bell pairs needed as a function of the working fidelity for different number of links $2^N$. (bottom) Plot of the distribution time as a function of the working fidelity for different number of links $2^N$ in units of $t_{\rm{segment}}$.}
\label{W095}
\end{figure}

\begin{figure}[htb]
\includegraphics[scale=0.45]{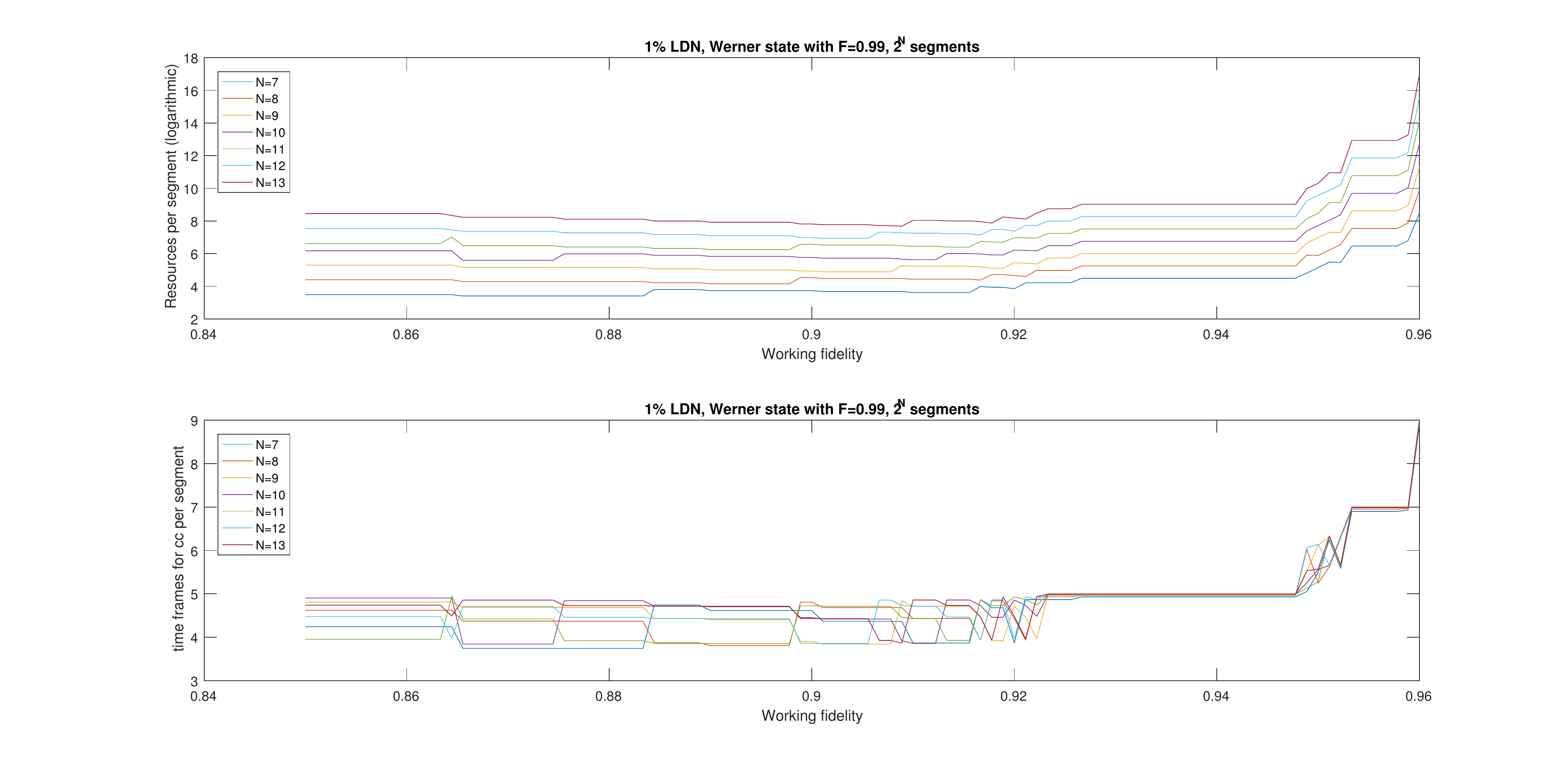}
\caption{(top) Plot of the number of elementary Bell pairs needed as a function of the working fidelity for different number of links $2^N$. (bottom) Plot of the distribution time as a function of the working fidelity for different number of links $2^N$ in units of $t_{\rm{segment}}$.}
\label{W099}
\end{figure}

\end{widetext}

\bibliographystyle{apsrev4-1}
\bibliography{repeater_hashing}

\end{document}